\documentclass[aps,pra,twocolumn,showpacs,groupedaddress,notitlepage,superscriptaddress,floatfix,nofootinbib]{revtex4-2}

\usepackage{braket}
\usepackage{amsmath}
\usepackage{amsfonts}
\usepackage{graphics}
\usepackage{graphicx}
\usepackage{mathtools}
\usepackage{hyperref}
\usepackage{subfigure}
\usepackage{placeins}
\usepackage[qm]{qcircuit}

\newcommand{\bes}{\begin{subequations}}
\newcommand{\ees}{\end{subequations}}
\newcommand{\beq}{\begin{equation}}
\newcommand{\eeq}{\end{equation}}

\hypersetup{
    colorlinks=true,   
    linkcolor=cyan,    
    citecolor=magenta, 
    filecolor=magenta, 
    breaklinks=true,
}

\usepackage{hyperref}
\hypersetup{
    colorlinks=true, 
    linkcolor=cyan,
    citecolor=magenta, 
    filecolor=magenta, 
    urlcolor=cyan,
    runcolor=cyan
}
\usepackage[capitalise]{cleveref}
\usepackage[ruled,vlined]{algorithm2e}

\begin{document}
\title{Universal Weakly Fault-Tolerant Quantum Computation via Code Switching in the [[8,3,2]] Code}

\author{Shixin Wu} 
\affiliation{Department of Electrical \& Computer  Engineering,  University of Southern California, Los Angeles, California 90089, USA}
\affiliation{Center for Quantum Information Science \& Technology, University of Southern California, Los Angeles, CA 90089, USA}
\author{Dawei Zhong}
\affiliation{Center for Quantum Information Science \& Technology, University of Southern California, Los Angeles, CA 90089, USA}
\affiliation{Department of Physics \& Astronomy, University of Southern California, Los Angeles, California 90089, USA}
\author{Todd A. Brun}
\affiliation{Department of Electrical \& Computer  Engineering,  University of Southern California, Los Angeles, California 90089, USA}
\affiliation{Center for Quantum Information Science \& Technology, University of Southern California, Los Angeles, CA 90089, USA}
\affiliation{Department of Physics \& Astronomy, University of Southern California, Los Angeles, California 90089, USA}
\affiliation{Department of Computer Science, University of Southern California, Los Angeles, California 90089, USA}
\author{Daniel A. Lidar}
\affiliation{Department of Electrical \& Computer  Engineering,  University of Southern California, Los Angeles, California 90089, USA}
\affiliation{Center for Quantum Information Science \& Technology, University of Southern California, Los Angeles, CA 90089, USA}
\affiliation{Department of Physics \& Astronomy, University of Southern California, Los Angeles, California 90089, USA}
\affiliation{Department of Chemistry, University of Southern California, Los Angeles, California 90089, USA}
\affiliation{Quantum Elements, Inc., 2829 Townsgate Road, Westlake Village, CA, 91361, USA}

\begin{abstract}
    Code-switching offers a route to universal, fault-tolerant quantum computation by circumventing the limitation implied by the Eastin-Knill theorem
    against a universal transversal gate set within a single quantum code.
    Here, we present a fault-tolerant code-switching protocol between two versions of the $[[8, 3, 2]]$ code. One version supports weakly fault-tolerant single-qubit Clifford gates, while the other supports a logical $\overline{\mathrm{CCZ}}$ gate via transversal $T/T^\dagger$ together with logical $\overline{\mathrm{CZ}}$, $\overline{\mathrm{CNOT}}$, and $\overline{\mathrm{SWAP}}$ gates. 
    Because both codes have distance 2, the protocol operates in a postselected, error-detecting regime: single faults lead to detectable outcomes, and accepted runs exhibit quadratic suppression of logical error rates. This yields a universal scheme for postselected fault-tolerant computation. We validate the protocol numerically through simulations of state preparation, code switching, and a three-logical-qubit implementation of Grover's search.
\end{abstract}

\maketitle
 
\section{Introduction}

Quantum error correction (QEC) protects quantum information from noise by encoding logical states into a code subspace or subsystem \cite{shor_scheme_1995, Steane:96a,Gottesman:1996fk,Calderbank:1997aa,Knill:1997kx,Knill:2000dq,Bacon:05,Kribs:2005:180501,poulin_stabilizer_2005,Kosut:2008lq,Gaitan:book,Lidar-Brun:book}. Beyond storage, quantum computation also requires controlled processing of encoded logical information. This is accomplished by applying quantum circuits that transform logical states while preserving the code. Moreover, these circuits must be \emph{fault-tolerant}: errors must not spread faster than the code's ability to correct them, or, for purely error-detecting (distance-$2$) codes, at least to detect them. When this condition is met for a suitable choice of code and for physical error rates below a threshold, the logical error rate can be suppressed to an arbitrarily low value for an arbitrarily long computation \cite{DiVincenzo:96,aharonov1997fault,Aliferis:05,reichardt_fault-tolerance_2005,Knill:05,Campbell:2017aa}. Very recently, the first empirical evidence of sub-threshold scaling has started to emerge~\cite{Acharya:2025aa,Eickbusch:2025aa,Lacroix:2025aa,Bluvstein:2026aa,vezvaee2025surfacecodescalingheavyhex}.

One of the simplest routes to fault tolerance is through \emph{transversal} gate implementations~\cite{Shor:96}. A transversal logical gate decomposes into physical gates that act independently on each qubit within a code block (and, for two-block gates, couple only corresponding qubits across blocks). This structure prevents a single physical error from spreading within a block and turning into a higher-weight error~\cite{eastin2009restrictions,Knill:05}. While scaling quantum computers to fully error-correcting fault tolerance remains challenging, a recent study demonstrated an approach to \emph{weak} fault tolerance in implementing the logical Clifford set on the high-rate $[[n,n-2,2]]$ family of codes without using any transversal logical gates \cite{gerhard2024weakly}. In this weakly fault-tolerant setting, circuits are designed so that any error produced by a single faulty gate or measurement is detectable by measuring ancillas used in their weakly fault-tolerant circuits that implement two-qubit entangling gates. Runs with nontrivial syndrome are discarded~\cite{Knill:05,gerhard2024weakly}. Although this approach is error-detecting rather than error-correcting, postselection yields a quadratic suppression of logical errors, consistent with the expected behavior of the $[[n,n-2,2]]$ family of codes, and does not require real-time decoding and feedback.

Encoded universal quantum computation requires a set of logical gates that can approximate any unitary transformation \cite{kitaev1997quantum}. A common minimal choice is $\{H,{\mathrm{CNOT}},T\}$, where $H$ and $T$ denote the Hadamard and $\pi/8$ gates, respectively~\cite{Boykin:99}. However, the Eastin-Knill theorem implies that no QEC code admits a universal set of \emph{transversal} logical gates \cite{eastin2009restrictions}. A widely used workaround is magic-state injection and distillation~\cite{Bravyi:2005aa}: noisy ``magic states'' are injected as non-Clifford resources, and distillation (using only Clifford operations) converts many noisy copies into fewer copies with higher fidelity. However, estimates indicate that realizing a single high-fidelity logical $T$ gate can require on the order of $10^3$ code cycles~\cite{Litinski2019magicstate}, which remains far beyond the reach of near-term platforms.

An alternative route to a fault-tolerant universal gate set is \emph{code switching} between codes that implement complementary logical gates via transversal physical gates~\cite{Anderson:2014aa}. Switching codes while preserving the logical information typically proceeds via \emph{gauge fixing}, i.e., measuring suitable gauge operators and projecting onto eigenstates so that the measured eigenoperators become stabilizer generators of the target code \cite{bombin2015gauge, kubica2015universal, quan2018fault,butt2024fault}.

Fault-tolerant code-switching protocols have been developed, for example, between the $[[7,1,3]]$ Steane code (the smallest 2D color code) and the $[[15,1,3]]$ Reed-Muller code (the smallest 3D color code) \cite{Anderson:2014aa,butt2024fault}. The same work also presented a protocol between the $[[7,1,3]]$ Steane code and a ``morphed'' $[[10,1,2]]$ code derived from the $[[15,1,3]]$ Reed-Muller code \cite{vasmer2022morphing}, reducing circuit depth and resource requirements, at the cost of reduced error-correction capability.

\begin{figure}[t]
    \centering
    \includegraphics[width=\linewidth]{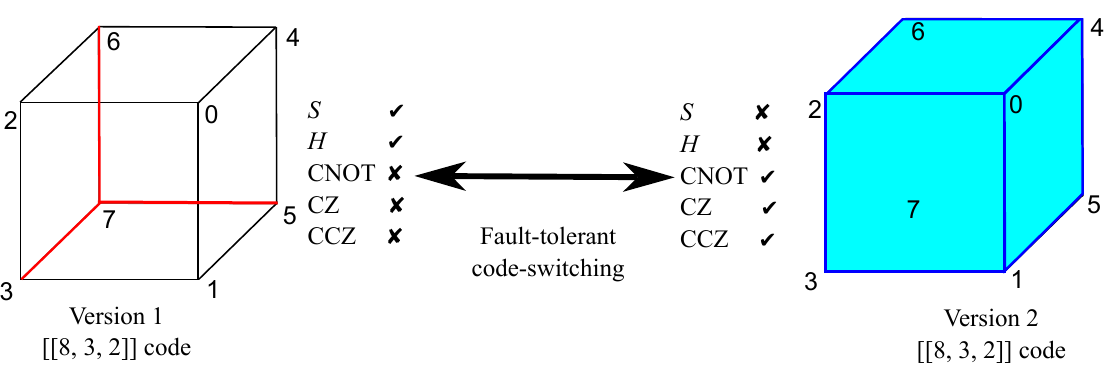}
    \caption{Summary of this work. The two versions of the $[[8,3,2]]$ code, each of which can be geometrically represented by a cube. The three red edges of the $[[8,3,2]]_1$ code represent the fixed weight-$2$ gauge $G^X$ operators, and the three blue faces of the $[[8,3,2]]_2$ code represent the fixed weight-$4$ gauge $G^Z$ operators.} 
    \label{fig:cs-1}
\end{figure}

In this work, we further reduce the required resources by code switching between two versions of the $[[8,3,2]]$ code. Because our codes have distance 2, we adopt fault tolerance in the error-detecting (postselected) sense. One version of the $[[8,3,2]]$ code is the ``smallest interesting color code,'' with fault-tolerant implementations of the logical CCZ, CZ, and CNOT gates \cite{campbell2016smallest, hangleiter2025fault, wang2024fault,honciuc2024implementing,bluvstein2024logical}. We call this the \emph{Version 2} $[[8,3,2]]$ code, because it lacks fault-tolerant implementations of logical single-qubit gates. The \emph{Version 1} $[[8,3,2]]$ code can instead use the weakly fault-tolerant circuits of Ref.~\cite{gerhard2024weakly} to implement the logical $H$, $S$, and $\sqrt{X}^\dagger$ gates. Together, the two codes provide an overcomplete universal set of logical gates while still achieving quadratic suppression of logical errors with postselection. This protocol is summarized in \cref{fig:cs-1}.

The remainder of the paper is organized as follows. We first review, in \cref{sec:2}, the weakly fault-tolerant logical gates available on the $[[n,n-2,2]]$ family. Then, in \cref{sec:3}, we introduce our two versions of the $[[8,3,2]]$ code.
In \cref{sec:4}, we present a fault-tolerant code-switching protocol between them. Next, in \cref{sec:5}, 
we summarize fault-tolerant state-preparation routines for logical states relevant to applications.
Finally, in \cref{sec:sim,sec:Grover}, we report numerical simulations of state preparation, code switching, and an encoded three-logical-qubit Grover search. The Appendix presents various technical details in support of the main text, including explicit circuits for fault-tolerant state-preparation.

\section{Fault-tolerant logical gates on $[[n, n-2, 2]]$ codes}
\label{sec:2}

This section summarizes the constructions in \cite{gerhard2024weakly} for implementing weakly fault-tolerant logical single-qubit Clifford gates ($H$, $S$, and $\sqrt{X}^\dagger$) on the $[[n,n-2,2]]$ code family with two-qubit entangling gates. Since these codes have distance $d=2$, our notion of fault tolerance is the $d=2$, postselected one: any single faulty gate during the circuit must either leave the data in the codespace or produce an error that is detected by the stabilizer (and ancilla) measurements, so that the run can be discarded.

The $[[n,n-2,2]]$ code has stabilizer generators
\beq
S_X = X^{\otimes n},\quad S_Z = Z^{\otimes n}.
\eeq
A convenient choice of logical Pauli operators is
\beq
\label{eq:canonical}
\overline{X}_j = X_jX_{n-2},\quad \overline{Z}_j = Z_jZ_{n-1}\quad (j=0,\ldots,n-3),
\eeq
with $\overline{Y}_j \equiv i\overline{X}_j\overline{Z}_j$ (global phases are irrelevant throughout).

Logical gates implemented \emph{transversally} are naturally compatible with weak fault tolerance: a single faulty location can create at most a weight-$1$ error in each code block, which remains detectable for a distance-$2$ code. By contrast, a faulty entangling gate that acts \emph{within} a block can 
produce a weight-$2$ error in the code block, potentially causing logical errors on a distance-$2$ code.
Such gates can still be used in a weakly fault-tolerant manner, but only when embedded in ancilla-assisted measurement gadgets that ensure any single fault produces a detectable syndrome.

It is useful to distinguish three conceptual levels:
\beq
\text{Logical level}  \rightarrow  \text{Physical level}  \rightarrow  \text{FT physical level}.
\eeq
At the \emph{logical level}, we denote ideal encoded operations with a straight overline (e.g., $\overline{H}_j$). At the \emph{physical level}, we write the corresponding compilation into elementary two-qubit rotations with a wavy overline (e.g., $\widetilde{XX}(i,j)$). These physical-level two-qubit entangling gates are not fault-tolerant on a distance-2 code and must be replaced by fault-tolerant gadgets at the \emph{FT physical level}; the gadgets we use are shown in \cref{app:A}. We write products of unitaries in the usual right-to-left circuit order (the rightmost factor acts first).

\subsection{Elementary Clifford rotations}
\label{sec:2A}

Both the physical-level compilation and the FT gadgets use the following Clifford rotations:
\bes
\begin{align}
U_{XX} &:= \frac{I+iXX}{\sqrt{2}},\quad
U_{YY} := \frac{I-iYY}{\sqrt{2}},\\
U_{ZZ} &:= \frac{I-iZZ}{\sqrt{2}},\quad
R_X := \frac{I-iX}{\sqrt{2}},
\end{align}
\ees
where $XX$, $YY$, and $ZZ$ denote products of Pauli operators on two qubits (e.g., $XX=X\otimes X$). We refer to these as the $XX$, $YY$, $ZZ$, and $R_X$ gates; note the opposite sign convention for the $XX$ gate. We use $\widetilde{XX}(i,j)$ (resp.\ $\widetilde{ZZ}(i,j)$) to denote applying $U_{XX}$ (resp.\ $U_{ZZ}$) to physical qubits $i$ and $j$.

For later reference, \cref{tab:stab_behavior} lists the conjugation action of the $XX$ and $ZZ$ rotations on a set of two-qubit Pauli operators used in the discussion.

\begin{table}
\centering
\begin{tabular}{|l|l|l|l|l|l|l|}
\hline
        & $XI$ & $ZI$ & $YX$  & $YZ$  & $XX$ & $ZZ$ \\ \hline
$XX$ gate & $XI$ & $YX$ & $-ZI$ & $YZ$  & $XX$ & $ZZ$ \\ \hline
$ZZ$ gate & $YZ$ & $ZI$ & $YX$  & $-XI$ & $XX$ & $ZZ$ \\ \hline
\end{tabular}
\caption{
Conjugation action of the $XX$ and $ZZ$ rotation gates on selected two-qubit Pauli operators (e.g., $XI$ denotes $X\otimes I$). Other Pauli mappings follow analogously.
}
\label{tab:stab_behavior}
\end{table}

\subsection{Logical level $\longrightarrow$ Physical level}
\label{sec:2B}

At the physical level, the logical single-qubit gates $\overline{S}_j$ (where $S$ is the phase gate), $\overline{\sqrt{X}_j^{\dagger}}$, and $\overline{H}_j$ can be expressed using $\widetilde{XX}$ and $\widetilde{ZZ}$ rotations involving the ``shared'' qubits $n-2$ and $n-1$:
\beq
\overline{S}_j = \widetilde{ZZ}(j,n-1),
\quad
\overline{\sqrt{X}_j^{\dagger}} = \widetilde{XX}(j,n-2),
\eeq
and
\beq
\overline{H}_j
= \widetilde{ZZ}(j,n-1)\cdot \widetilde{XX}(j,n-2)\cdot \widetilde{ZZ}(j,n-1)\cdot Y_jX_{n-2}Z_{n-1}.
\eeq
The final factor 
$Y_jX_{n-2}Z_{n-1}$
is a Pauli byproduct correction (not merely a global phase). The sequence 
$\widetilde{ZZ}\cdot \widetilde{XX}\cdot \widetilde{ZZ}$
implements $\overline{H}_j$ up to a known logical $\overline{Y}_j$ correction, and
with our choice of logical operators,
\beq
\overline{Y}_j \equiv Y_jX_{n-2}Z_{n-1}
\eeq
up to an irrelevant phase. 

The identities above follow from \cref{tab:stab_behavior} together with the standard single-qubit Clifford conjugation rules
\beq
\begin{aligned}
&SXS^\dagger=Y, \quad SYS^\dagger=-X, \quad SZS^\dagger=Z,\\
&\sqrt{X}^{\dagger}X\sqrt{X}=X,\quad
\sqrt{X}^{\dagger}Y\sqrt{X}=-Z,\quad
\sqrt{X}^{\dagger}Z\sqrt{X}=Y,\\
&HXH=Z,\quad HYH=-Y,\quad HZH=X.
\end{aligned}
\eeq
For example, $\widetilde{ZZ}(j,n-1)$ leaves $\overline{Z}_j$ invariant and maps $\overline{X}_j\mapsto \overline{Y}_j$ (up to phase), matching the action of $\overline{S}_j$. Similarly, $\widetilde{XX}(j,n-2)$ leaves $\overline{X}_j$ invariant and maps $\overline{Z}_j\mapsto \overline{Y}_j$, matching the action of $\overline{\sqrt{X}_j^{\dagger}}$.

\subsection{Physical level $\longrightarrow$ FT physical level}
\label{sec:2C}

The two-qubit entangling gates $\widetilde{XX}$ and $\widetilde{ZZ}$ are not automatically fault-tolerant on a distance-2 code, because they may produce weight-$2$ errors on the data qubits that are logical errors in most cases. Because the probabilities for two-qubit entangling gates to produce weight-$1$ and weight-$2$ errors are both on the order of $O(p)$, where $p$ is the error probability of the gate, logical errors cannot be quadratically suppressed. Ref.~\cite{gerhard2024weakly} replaces each $\widetilde{XX}(j,n-2)$ and $\widetilde{ZZ}(j,n-1)$ by an ancilla-assisted, verification/postselection gadget that is fault-tolerant in the error-detecting sense where all two-qubit entangling gates act on a data qubit and an ancilla and none on two data qubits. We reproduce these gadgets in \cref{app:A} (\cref{fig:FTXXZZ}) and use them as primitives in the remainder of this paper.

\section{Two versions of the $[[8,3,2]]$ code}
\label{sec:3}
\begin{figure}[t]
    \centering
    \includegraphics[width=\linewidth]{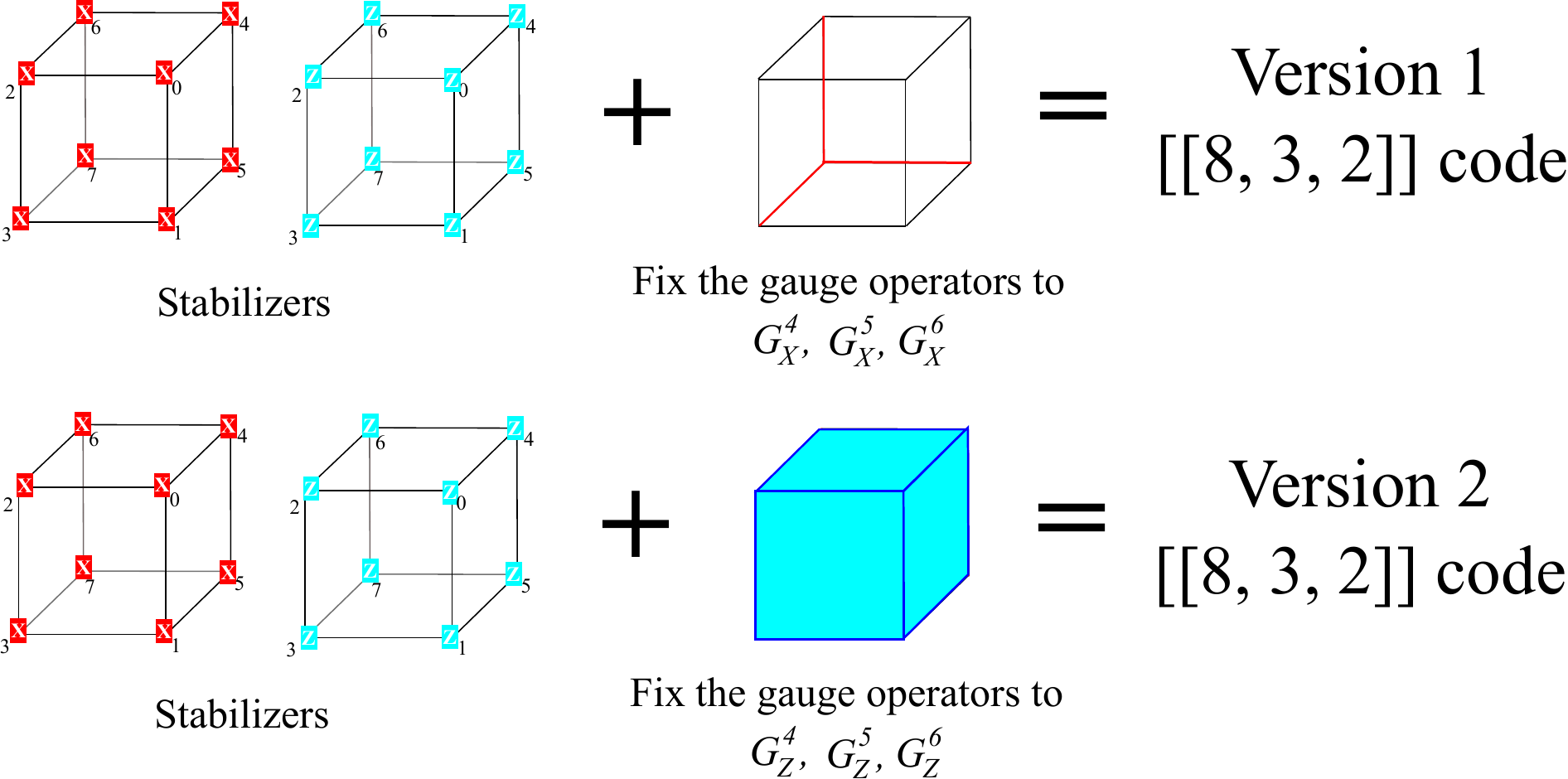}
    \caption{Obtaining the two versions of the $[[8,3,2]]$ code by fixing the gauge operators of the $[[8,3,3,2]]$ subsystem code. Together with the transversal $X^{\otimes 8}$ and $Z^{\otimes 8}$ stabilizers of the $[[8,3,3,2]]$ subsystem code, fixing the gauge operators gives the different stabilizers that define the two stabilizer $[[8,3,2]]$ codes.}
    \label{fig:cs-2}
\end{figure}

In this section, we discuss two versions of the $[[8,3,2]]$ stabilizer code: the Version 1 $[[8,3,2]]$ code ($[[8,3,2]]_1$) and the Version 2 $[[8,3,2]]$ code ($[[8,3,2]]_2$), as well as their construction from the $[[8,6,2]]$ code. The constructions of the two versions are summarized in \cref{fig:cs-2}.

\subsection{Subsystem $[[8,3,3,2]]$ code}
The subsystem $[[8, 3, 3, 2]]$ code~\cite{Singkanipa2025familiesofdd} is constructed from the $n=8$ $[[n, n-2, 2]]$ stabilizer code by ``gauging out'' three of its six logical operators, i.e., not using them to store logical data and not requiring that they be protected. The stabilizers of the $[[8, 6, 2]]$ code are $X^{\otimes 8}$ and $Z^{\otimes 8}$. We then choose the following logical operators, which differ from the canonical all-weight-$2$ choice in \cref{eq:canonical}. This choice is crucial for the code-switching protocol to function, and corresponds to changing the basis of the codespace.
\begin{equation}
\label{eq:original862}
    \begin{split}
        \overline{X}_1&=X_0X_1X_2X_3,\ \overline{Z}_1=Z_0Z_4 \\
        \overline{X}_2&=X_0X_1X_4X_5,\ \overline{Z}_2=Z_0Z_2 \\
        \overline{X}_3&=X_0X_2X_4X_6,\ \overline{Z}_3=Z_0Z_1 \\
        G^X_4&=X_3X_7, \ G^Z_4=Z_0Z_1Z_2Z_3 \\
        G^X_5&=X_5X_7, \ G^Z_5=Z_0Z_1Z_4Z_5 \\
        G^X_6&=X_6X_7, \ G^Z_6=Z_0Z_2Z_4Z_6 \\
    \end{split}
\end{equation}
Gauging out the last three logical qubits gives us the $[[8, 3, 3, 2]]$ subsystem code. We refer to the last three logical qubits as ``gauge qubits'' and their corresponding logical operators $G_{i=4,5,6}^{X/Z}$ as ``gauge operators.''

\subsection{The $[[8,3,2]]_1$ stabilizer code}
Version 1 of the $[[8, 3, 2]]$ stabilizer code is obtained by fixing the gauge qubits of the $[[8, 3, 3, 2]]$ code into the logical $\ket{\overline{+}}$ state. This means the logical $X$ operators of the three gauge qubits now become stabilizers of the code. Thus, the new set of stabilizer generators is
\begin{equation}
\label{eq:v1_stabilizers}
    X^{\otimes 8}, Z^{\otimes 8},  X_3X_7, X_5X_7, X_6X_7. 
\end{equation}
The additional $X$ stabilizer generators reduce the first three logical operators to weight-$2$:
\begin{equation}
\label{eq:v1_logical_x}
    \begin{split}
        \overline{X}_1&=X_0X_1X_2X_3 \cdot X^{\otimes 8}\cdot X_5X_7=X_4X_6,  \\
        \overline{X}_2&=X_0X_1X_4X_5 \cdot X^{\otimes 8}\cdot X_3X_7=X_2X_6,  \\
        \overline{X}_3&=X_0X_2X_4X_6 \cdot X^{\otimes 8}\cdot X_3X_5X_6X_7 = X_1X_6.\\
    \end{split}
\end{equation}
Its logical $Z$ operators remain unchanged:
\begin{equation}
    \overline{Z}_1=Z_0Z_4,\ \overline{Z}_2=Z_0Z_2,\ \overline{Z}_3=Z_0Z_1.
\end{equation}
Note that the logical operators are similar to those used in \cite{gerhard2024weakly}, in that all of them are weight-$2$ [instead of a mix of weight-$4$ and weight-$2$ as in \cref{eq:original862}] and, 
 within each Pauli type, the same shared qubit is used for all three logical qubits: qubit $6$ for the logical $\overline{X}$ operators and qubit $0$ for the logical $\overline{Z}$ operators.
 As a result, we can use the same construction as in \cref{sec:2B} to implement the logical $\overline{H}$, $\overline{S}$, and $\overline{\sqrt{X}^\dagger}$ gates
for the Version 1 $[[8, 3, 2]]$ code:
\begin{equation}
    \begin{aligned}
    \label{eq:hadamard}
        \overline{H}_1 &= \widetilde{ZZ}(4,0) \cdot \widetilde{XX}(4,6) \cdot \widetilde{ZZ}(4,0) \cdot Y_4X_{6}Z_0, \\
        \overline{H}_2 &= \widetilde{ZZ}(2,0) \cdot \widetilde{XX}(2,6) \cdot \widetilde{ZZ}(2,0) \cdot Y_2X_{6}Z_0, \\
        \overline{H}_3 &= \widetilde{ZZ}(1,0) \cdot \widetilde{XX}(1,6) \cdot \widetilde{ZZ}(1,0) \cdot Y_1X_{6}Z_0. \\
        \overline{S}_1 &= \widetilde{ZZ}(4,0),\quad  \overline{\sqrt{X}_1^\dagger} = \widetilde{XX}(4,6), \\
        \overline{S}_2 &= \widetilde{ZZ}(2,0),\quad  \overline{\sqrt{X}_2^\dagger} = \widetilde{XX}(2,6), \\
        \overline{S}_3 &= \widetilde{ZZ}(1,0),\quad  \overline{\sqrt{X}_3^\dagger} = \widetilde{XX}(1,6). \\
    \end{aligned}
\end{equation}
Then, we use the fault-tolerant implementation of \cref{fig:FTXXZZ} to replace $\widetilde{XX}$ and $\widetilde{ZZ}$. 
Finally, logical SWAPs among the three encoded qubits are induced by the corresponding permutations of the physical qubits $\{1,2,4\}$, since these are the qubits that distinguish the three logical-qubit labels. 
On hardware platforms with dynamically reconfigurable or movable qubits, such as reconfigurable neutral-atom arrays and shuttling trapped-ion architectures, these operations can often be absorbed into a software-level remapping of qubit identities rather than implemented as explicit logical gate sequences \cite{Tan2024compilingquantum,Bluvstein2022coherent,Pino2021qccd}. On architectures without that flexibility, the same permutations must instead be compiled into explicit routing operations, typically SWAP networks on devices with restricted connectivity \cite{Li2019sabre,Cowtan2019routing}.

Although a fault-tolerant implementation of the logical intra-block CNOT gate was also proposed in Ref.~\cite{gerhard2024weakly}, we opt not to use it in the Version 1 $[[8,3,2]]$ code due to its large circuit depth. Rather, CNOTs can be easily implemented in Version 2, as we show next.

\subsection{The $[[8,3,2]]_2$ stabilizer code}
Version 2 of the $[[8, 3, 2]]$ stabilizer code is obtained by fixing the gauge qubits of the $[[8, 3, 3, 2]]$ code in the logical $\ket{\overline{0}}$ state. This means that the logical $Z$ operators of the gauge qubits now become stabilizer generators for the code. Thus, the new set of stabilizer generators is
\beq
    X^{\otimes 8},Z^{\otimes 8}, Z_0Z_1Z_2Z_3, Z_0Z_1Z_4Z_5, Z_0Z_2Z_4Z_6.
\eeq

This recovers the $[[8,3,2]]$ code (version 2) that allows a fault-tolerant logical $\mathrm{CCZ}$ gate with transversal $T$ and $T^\dagger$ gates \cite{campbell2016smallest}:
\beq
    \overline{\mathrm{CCZ}} = T_0T_1^\dagger T_2^\dagger T_3T_4^\dagger T_5T_6T_7^\dagger.
\eeq
In addition, this code also allows fault-tolerant implementations of the logical CZ, SWAP, intra-block CNOT gates, and inter-block CNOT gates \cite{wang2024fault,honciuc2024implementing,bluvstein2024logical, hangleiter2025fault}.
The logical CZ gates are implemented with phase gates:
\begin{equation}
    \begin{aligned}
        &\overline{\mathrm{CZ}}(1,2) = S_0S_2^\dagger S_4^\dagger S_6,\\
        &\overline{\mathrm{CZ}}(1,3) = S_0S_1^\dagger S_4^\dagger S_5,\\
        &\overline{\mathrm{CZ}}(2,3) = S_0S_1^\dagger S_2^\dagger S_3.
    \end{aligned}
\end{equation}

The logical SWAP gates are implemented by cyclic-permuting the physical qubits: 
\begin{equation}
\small
    \begin{split}
        \overline{\mathrm{SWAP}(1,2)}&: 0 \to 4 \to 6 \to 2 \to 0, \ 1 \to 5 \to 7 \to 3 \to 1,\\
        \overline{\mathrm{SWAP}(1,3)}&: 2 \to 6 \to 7 \to 3 \to 2, \ 0 \to 4 \to 5 \to 1 \to 0, \\
        \overline{\mathrm{SWAP}(2,3)}&: 0 \to 2 \to 3 \to 1 \to 0, \ 4 \to 6 \to 7 \to 5 \to 4. \\
    \end{split}
\end{equation}
Similarly, these physical SWAP operations can be treated either as qubit permutations or as explicit qubit motions, depending on the layout of different architectures. 

Likewise, the following intra-block logical CNOT gates are induced by physical-qubit permutations:
\begin{equation}
\small
    \begin{split}
        \overline{\mathrm{CNOT}(1,2)}=0 \leftrightarrow 4,1 \leftrightarrow 5,&\quad \overline{\mathrm{CNOT}(2,1)}=0 \leftrightarrow 2,1 \leftrightarrow3 \\
        \overline{\mathrm{CNOT}(1,3)}=0 \leftrightarrow 4,2 \leftrightarrow 6,&\quad \overline{\mathrm{CNOT}(3,1)}=0 \leftrightarrow 1,2 \leftrightarrow3 \\
        \overline{\mathrm{CNOT}(2,3)}=0 \leftrightarrow 2,4 \leftrightarrow 6,&\quad \overline{\mathrm{CNOT}(3,2)}=0 \leftrightarrow 1,4 \leftrightarrow5 \\
    \end{split}
\end{equation}

\subsection{Block-scalability}
We consider scaling up with blocks of the $[[8,3,2]]$ code, which requires performing entangling gates between logical qubits in different blocks. 

A parallel inter-block logical $\overline{\mathrm{CNOT}}$ gate means that the transversal physical CNOT gates are applied between corresponding qubits of two code blocks, so that the three corresponding logical qubits undergo logical $\overline{\mathrm{CNOT}}$ gates in parallel. When the two blocks are in the same version of the $[[8,3,2]]$ code, this transversal CNOT preserves the CSS stabilizer structure and therefore defines a valid encoded operation~\cite{Gottesman:97b}. For mixed versions, the situation is asymmetric. If the control block is $[[8,3,2]]_1$ and the target block is $[[8,3,2]]_2$, the fixed gauge-$X$ operators $G^X_{4,5,6}$ of the control are copied to the target under transversal CNOT, so the target block is taken out of the $[[8,3,2]]_2$ codespace. In the reverse direction, with $[[8,3,2]]_2$ as control and $[[8,3,2]]_1$ as target, no such fixed gauge-$X$ operators are copied, so the codespace is preserved. Therefore a parallel inter-block logical $\overline{\mathrm{CNOT}}$ is available between mixed-version blocks only in the direction $[[8,3,2]]_2 \to [[8,3,2]]_1$.

\begin{figure}
    \centering
    \includegraphics[width=0.8\linewidth]{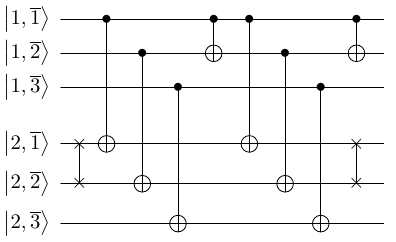}
    \caption{The logical circuit for the single-pair inter-block 
    $\overline{\mathrm{CNOT}}$ from logical qubit $\overline{1}$ of block $1$ to logical qubit $\overline{1}$ of block $2$ \cite{hangleiter2025fault}. If the two logical $\overline{\mathrm{SWAP}}$ gates are omitted, the circuit instead implements $\overline{\mathrm{CNOT}}(\ket{1,\overline{1}},\ket{2,\overline{2}})$; the SWAPs simply swap the target from $\ket{2,\overline{2}}$ to $\ket{2,\overline{1}}$. Variants of this construction are given in \cref{app:C}.}
    \label{fig:interblock-CNOT}
\end{figure}

The logical circuit for a single-pair inter-block $\overline{\mathrm{CNOT}}$ is shown in \cref{fig:interblock-CNOT} \cite{hangleiter2025fault}. It uses only logical $\overline{\mathrm{SWAP}}$ gates, parallel inter-block logical $\overline{\mathrm{CNOT}}$ gates, and intra-block logical $\overline{\mathrm{CNOT}}$ gates. 
Consequently, such a single-pair inter-block logical $\overline{\mathrm{CNOT}}$ is available whenever those three ingredients are available. They are always available between two Version 1 blocks or between two Version 2 blocks in principle, but in practice, the Version 2 realization uses much simpler circuits to implement the intra-block logical $\overline{\mathrm{CNOT}}$ gates. 
Between mixed-version blocks, the same asymmetry as above remains, so a single-pair inter-block logical $\overline{\mathrm{CNOT}}$ is possible from $[[8,3,2]]_2$ to $[[8,3,2]]_1$ but not in the reverse direction. 

An inter-block logical $\overline{\mathrm{CCZ}}$ gate (with participating logical
qubits initially residing in different blocks) can be implemented 
after first switching the relevant blocks into a compatible version (typically Version 2). One then uses inter-block logical $\overline{\mathrm{SWAP}}$ operations implemented with inter-block $\overline{\mathrm{CNOT}}$ gates
to bring the three participating
logical qubits into a common block, 
applies the intra-block logical $\overline{\mathrm{CCZ}}$ gate there, and finally reverses the SWAPs to return the logical qubits back to their original blocks.

\section{Code-switching between Two Versions of the $[[8,3,2]]$ code}
\label{sec:4}

The two versions of the $[[8,3,2]]$ code are different gauge fixings of the parent $[[8,3,3,2]]$ subsystem code. A direct switch by applying Hadamard gates to the three gauge qubits would exchange the $X$-type and $Z$-type gauge operators, but we do not know a fault-tolerant implementation of those gauge-qubit Hadamards, because the corresponding gauge operators have different weights in the two versions. We therefore switch between the two codes by measuring the gauge operators of the target version and, if necessary, applying gauge corrections that return all three gauge outcomes to $+1$.

\subsection{Switching from Version 2 to 1}

To switch from Version 2 to Version 1, we measure $G^X_4$, $G^X_5$, and $G^X_6$. These measurements project the gauge qubits onto eigenspaces of the $X$-type gauge operators. For each outcome $M_i^X=-1$, we apply the corresponding operator $G_i^Z$; we refer to this as a \emph{gauge correction}.

For fault tolerance, we also measure a complementary check $X_0X_1X_2X_4$ that satisfies
\beq
G^X_4 G^X_5 G^X_6 \cdot X_0X_1X_2X_4 = X^{\otimes 8}.
\eeq
Hence, the product of the three gauge outcomes and the outcome of $X_0X_1X_2X_4$ is the pre-correction $X^{\otimes 8}$ syndrome. If this parity is $-1$, we discard the run before applying any gauge correction. We also verify the target-code stabilizer $X^{\otimes 8}$. The gauge outcomes alone determine the correction, but the complementary check and the $X^{\otimes 8}$ stabilizer verification are needed when the switched block will subsequently be used for logical gates.

\subsection{Switching from Version 1 to 2}

The reverse direction is analogous. We measure $G^Z_4$, $G^Z_5$, and $G^Z_6$, and for each outcome $M_i^Z=-1$ we apply $G_i^X$. We also measure a complementary check $Z_1Z_2Z_4Z_7$ chosen so that
\begin{equation}
G^Z_4 G^Z_5 G^Z_6 \cdot Z_1Z_2Z_4Z_7 = Z^{\otimes 8}.
\end{equation}
Therefore, the product of the four measurement outcomes gives the pre-correction $Z^{\otimes 8}$ syndrome. If this parity is $-1$, we discard the run before applying any gauge correction. We also verify the target-code stabilizer $Z^{\otimes 8}$. Again, the gauge outcomes determine the correction, while the complementary check and the $Z^{\otimes 8}$ stabilizer verification ensure that the switched block is in the correct sector before any Version~2 logical gate is applied.

\subsection{Remarks on fault tolerance}

Fault-tolerant code switching must do more than preserve the detectability of a single data error. It must also place the block in the correct gauge and stabilizer sector of the target code before any subsequent logical gate is applied.

This matters in both versions of the code. In the $[[8,3,2]]_2$ code, the transversal $T/T^\dagger$ pattern implements the logical $\overline{\mathrm{CCZ}}$ gate only on the $+1$ eigenspace of the $Z$-type stabilizers, including $G^Z_4$, $G^Z_5$, and $G^Z_6$. In the $[[8,3,2]]_1$ code, the convenient weight-$2$ representatives of the logical $\overline{X}$ operators differ from the original weight-$4$ representatives by products of $G^X_4$, $G^X_5$, and $G^X_6$. A wrong gauge sign can therefore change the action of later logical gates, even if a later full target-code stabilizer measurement round, including the fixed gauge operators, would eventually reject the state.

The complementary check and the $X^{\otimes 8}$/$Z^{\otimes 8}$ stabilizer verification play different roles. The complementary check detects any single fault that flips the relevant $X^{\otimes 8}$/$Z^{\otimes 8}$ stabilizer syndrome before a gauge correction is applied, including a single ancilla fault during a gauge measurement. The $X^{\otimes 8}$/$Z^{\otimes 8}$ stabilizer verification detects faults that occur between successive gauge measurements and can therefore evade the pre-correction parity while still leading to an incorrect gauge correction. For example, in the Version-1-to-Version-2 protocol, an $X_0$ fault after the $G^Z_4$ measurement changes the remaining gauge outcomes to $(+1,-1,-1)$ while leaving $Z_1Z_2Z_4Z_7$ unchanged. The pre-correction parity, therefore, remains $+1$, so one would incorrectly apply $G^X_5$ and $G^X_6$; the subsequent $Z^{\otimes 8}$ verification would then reject the run, as it detects the $X_0$ error.

The two protocols are summarized in Algorithms \ref{alg: 1to2} and \ref{alg: 2to1} below, where $\mathrm{Meas}(P)$ denotes the eigenvalue $\pm 1$ returned by measuring the Pauli operator $P$. 

\begin{algorithm}
\caption{Fault-tolerant code switching from Version 1 to Version 2}
\label{alg: 1to2}

Measure the gauge operators\\
$M^Z_4 \gets \text{Meas}(G^Z_4)$,
$M^Z_5 \gets \text{Meas}(G^Z_5)$,
$M^Z_6 \gets \text{Meas}(G^Z_6)$\;

Measure the complementary check\\
$M^Z_7 \gets \text{Meas}(Z_1 Z_2 Z_4 Z_7)$\;

Verify the $Z^{\otimes 8}$ stabilizer\\
$M^Z_8 \gets \text{Meas}(Z^{\otimes 8})$\;

\If{$\prod_{i=4,5,6,7} M^Z_i = +1$ and $M^Z_8=+1$}{
    \ForEach{$i \in \{4,5,6\}$ such that $M_i^Z=-1$}{
        Apply $G_i^X$
    }
}
\Else{
    Discard current run\;
}
\end{algorithm}

\begin{algorithm}
\caption{Fault-tolerant code switching from Version 2 to Version 1}
\label{alg: 2to1}

Measure the gauge operators\\
$M^X_4 \gets \text{Meas}(G^X_4)$,
$M^X_5 \gets \text{Meas}(G^X_5)$,
$M^X_6 \gets \text{Meas}(G^X_6)$\;

Measure the complementary check\\
$M^X_7 \gets \text{Meas}(X_0 X_1 X_2 X_4)$\;

Verify the $X^{\otimes 8}$ stabilizer\\
$M^X_8 \gets \text{Meas}(X^{\otimes 8})$\;

\If{$\prod_{i=4,5,6,7} M^X_i = +1$ and $M^X_8=+1$}{
    \ForEach{$i \in \{4,5,6\}$ such that $M_i^X=-1$}{
        Apply $G_i^Z$
    }
}
\Else{
    Discard current run\;
}
\end{algorithm}

\begin{figure}[h]
    \centering
    \subfigure[]{
        \resizebox{0.20\textwidth}{!}{
        \Qcircuit @C=0.9em @R=1.1em {
         & \ctrl{4} & \qw      & \qw      & \qw      & \qw      & \qw      & \qw \\
         & \qw      & \qw      & \ctrl{3} & \qw      & \qw      & \qw      & \qw \\
         & \qw      & \qw      & \qw      & \ctrl{2} & \qw      & \qw      & \qw \\
         & \qw      & \qw      & \qw      & \qw      & \qw      & \ctrl{1} & \qw \\
        \lstick{\ket{0}} 
         & \targ    & \targ    & \targ    & \targ    & \targ    & \targ    & \rstick{Z} \qw \\
        \lstick{\ket{+}} 
         & \qw & \ctrl{-1}& \qw & \qw      & \ctrl{-1}& \qw      & \rstick{X} \qw
        }
        }
    }
    \hspace{2em}
    \subfigure[]{
        \resizebox{0.1\textwidth}{!}{
        \Qcircuit @C=1.0em @R=1.2em {
         & \targ    & \qw & \qw \\
         & \qw      & \targ &\qw \\
        \lstick{\ket{+}} & \ctrl{-2} & \ctrl{-1} &\rstick{X}\qw
        }
        }
    }
    \caption{
    (a) Flag-based measurement of a weight-$4$ $Z$-type gauge operator used in switching from Version 1 to Version 2. The upper four wires are the data qubits in the support of the measured operator, the fifth wire is the ancilla measured in the $Z$ basis to decide whether $G^X$ corrections are necessary, and the last wire is a flag ancilla measured in the $X$ basis. A triggered flag identifies single-fault events that can propagate to weight-$2$ $Z$ errors on the data qubits. (b) Fault-tolerant measurement of a weight-$2$ $X$-type gauge operator used in switching from Version 2 to Version 1. The upper two wires are the data qubits in the support of the measured operator, the third wire is the ancilla measured in the $X$ basis to decide whether $G^Z$ corrections are necessary. Because the measured operator has weight 2, no flag ancilla is required.}
    \label{fig:measure} 
\end{figure}
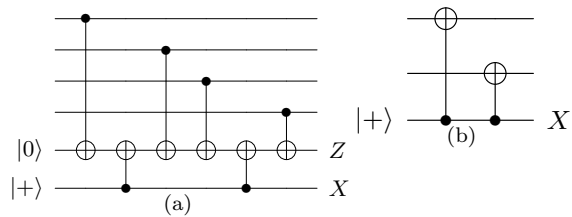

As in syndrome extraction, weight-$4$ gauge measurements require flag qubits \cite{chao2018quantum} to detect single ancilla faults that can propagate to weight-$2$ data errors. Therefore, switching from Version 1 to Version 2 requires flag qubits, because the measured $G^Z$ operators have weight $4$; see \cref{fig:measure}. Switching from Version 2 to Version 1 does not, because the measured $G^X$ operators have weight $2$ and a single ancilla fault cannot propagate to a higher weight.

\section{Fault-tolerant encoding circuits}
\label{sec:5}

We require fault-tolerant preparation routines for several logical input states in both versions of the $[[8,3,2]]$ code. We use $\ket{\overline{+^20^1}}$ and $\ket{\overline{+^10^2}}$ to denote mixed logical-basis states, specifying only the numbers of $\ket{\overline{+}}$ and $\ket{\overline{0}}$ states and not their ordering. The detailed stabilizer reductions, circuit constructions, and explicit encoding circuits are given in \cref{app:B}; here we summarize the main results.

Most of the relevant encoded states reduce, after Gaussian elimination of their stabilizer generators, to standard entangled resource states (GHZ states, dual GHZ states, Bell pairs, or tensor products thereof), so their preparation can be built from standard fault-tolerant GHZ/Bell-state gadgets. In particular, both versions admit fault-tolerant preparation of $\ket{\overline{000}}$ and Version~1 admits a direct fault-tolerant preparation of  $\ket{\overline{{++}+}}$. Without additional gauge fixing, Version~1 naturally supports states of the form $\ket{\overline{+^20^1}}$, while Version~2 naturally supports states of the form $\ket{\overline{+^10^2}}$. 
For Version~2, we use the preparation circuit of \cite{honciuc2024implementing}; in the noisy-initialization Grover benchmark, we augment it with additional flagged $G^Z_5$ and $G^Z_6$ checks to monitor harmful gauge-sector faults.

When the opposite mixed-basis pairing is required, namely, $\ket{\overline{+^20^1}}$ in Version 2 or $\ket{\overline{+^10^2}}$ in Version 1, we do not perform a full code-switching. Instead, we prepare a partially gauge-fixed intermediate state and measure the remaining single gauge operator, along with its complementary check, before immediately applying the stabilizer and gauge checks. This achieves the desired encoded state with lower overhead than a full switching subroutine while still ensuring that the block is in the correct gauge and stabilizer sector for subsequent logical operations.

\Cref{app:B} collects the explicit initialization circuits drawn from these constructions. Their simulated performance is discussed in the next section.

\section{\label{sec:sim} Simulation}
In this section, we use numerical simulations to evaluate the logical error rates of our state-preparation circuits and code-switching protocol in the presence of faulty gates. A faulty gate is modeled as $\widetilde{\mathcal{U}} = \mathcal{N} \circ \mathcal{U}$, where $\mathcal{U}(\rho) = U\rho U^{\dagger}$ is a noise-free unitary channel and $\mathcal{N}$ is the noise channel associated with the gate. In our simulations, we adopt the depolarizing noise channel for all two-qubit gates with physical error rate $p$ and for all single-qubit gates with physical error rate $q$. For simplicity, we neglect noise from physical qubit initialization, measurements, feedback, and corrections. The single-qubit and two-qubit depolarizing channels are given by
\begin{align}
    \mathcal{N}_1(\rho) =& (1-q)\rho + \frac{q}{3}\sum_{j} P_j \rho P_j,\\
    \mathcal{N}_2(\rho) =& (1-p)\rho + \frac{p}{15}\sum_{j} P_j \rho P_j.
\end{align}
where $\{P_j\}$ are all possible single/two-qubit Pauli operators except the identity. Thus, one of the possible single-qubit Pauli errors is applied with probability $q/3$ after each single-qubit gate, and one of the possible two-qubit Pauli errors is applied with probability $p/15$ after each two-qubit gate. 
The state-preparation and code-switching benchmarks
are performed using the Python package \texttt{PECOS}~\cite{PECOSPackage}, a tool designed for efficient Monte-Carlo simulations of quantum Clifford circuits.
The precise set of noisy operations is specified in each subsection; when single-qubit gate noise is included, we take $q=p/10$.

\subsection{State preparation}
\label{sec:state-prep}

\begin{figure*}
    \centering
    \includegraphics[width=0.9\textwidth]{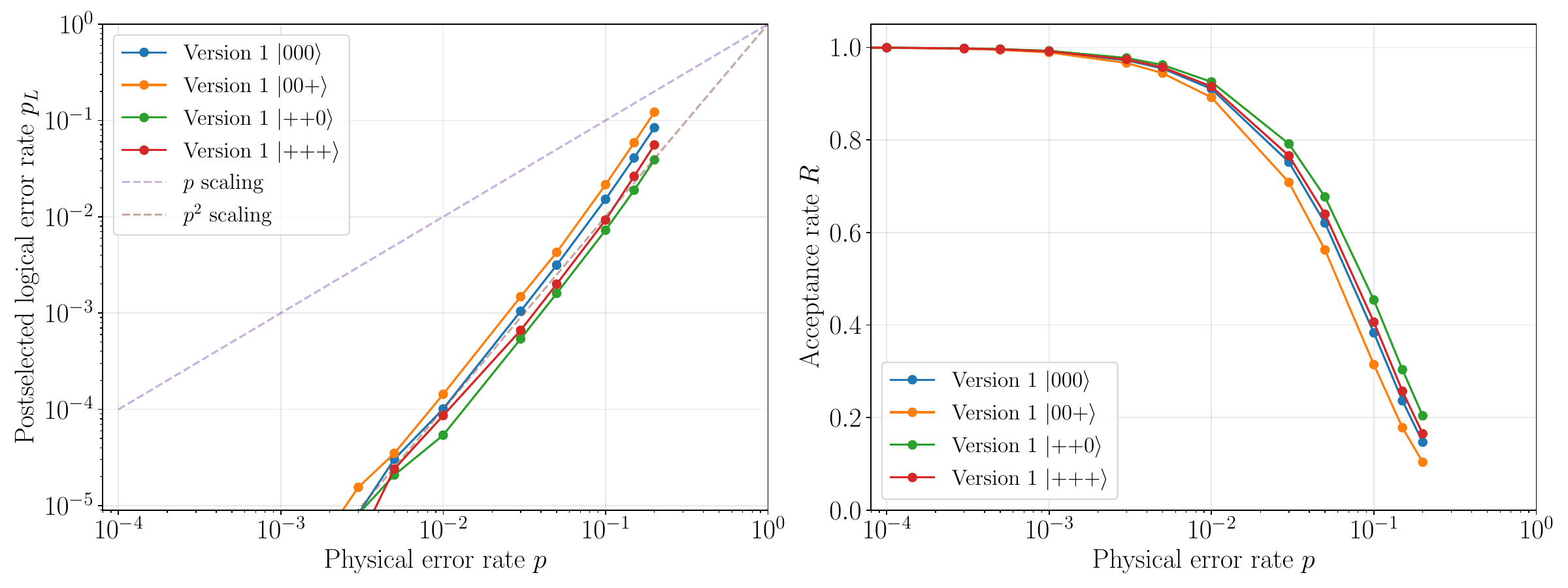}
   \includegraphics[width=0.9\textwidth]{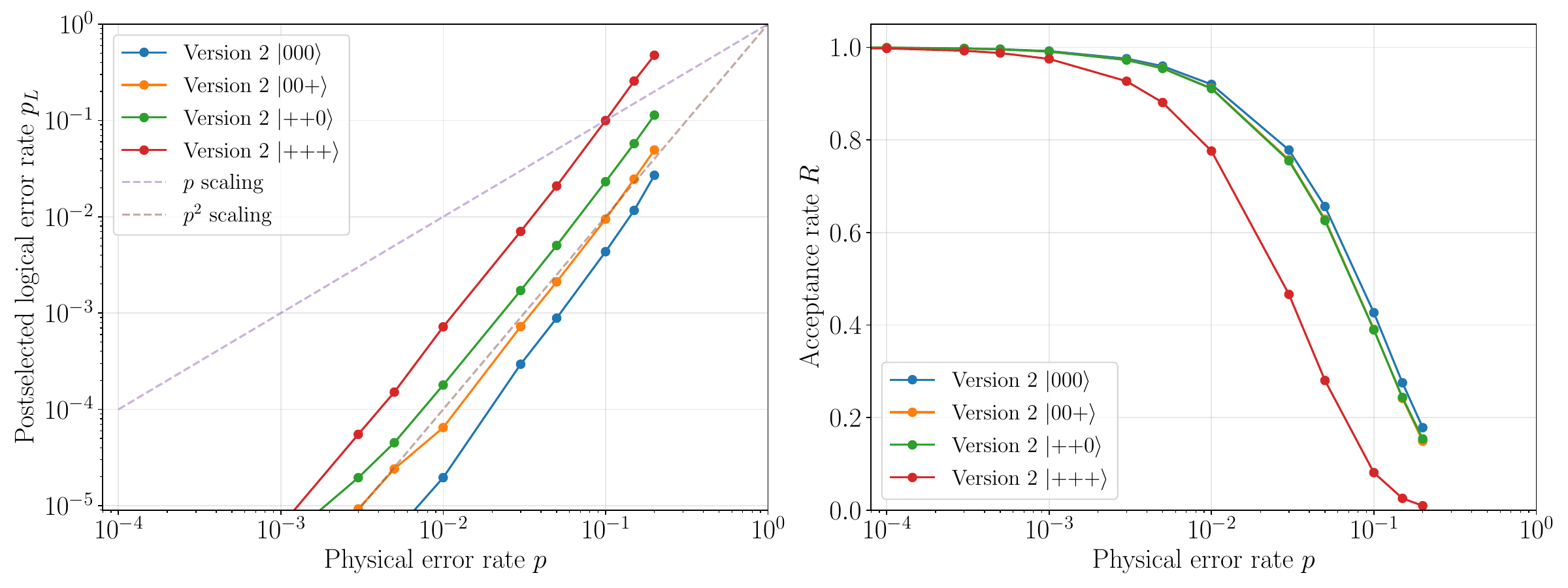}
    \caption{Numerical simulation of fault-tolerant state preparation for the $[[8,3,2]]$ code; see \cref{app:B} for the corresponding circuits. Top row: Version 1. Bottom row: Version 2. Left panels: postselected logical error rate $p_L$. Right panels: acceptance rate $R$. In these simulations, depolarizing noise with physical error rate $p$ is applied to every CNOT gate. In the low-noise regime, the observed behavior is consistent with weak fault tolerance: $p_L$ scales approximately quadratically with $p$, while $R$ decreases as detectable faults are removed by postselection.}
    \label{fig:sim_v12_init}
\end{figure*}

In the presence of noise, each state-preparation run falls into one of three categories: successful runs, discarded runs, and failed runs. 
A run is discarded whenever postselection detects a fault, either through a nontrivial flag outcome or through the final noiseless stabilizer and gauge measurements. Among the accepted runs, the preparation is counted as successful if the output matches the target logical state up to stabilizers, and as a failure otherwise. 
For example, preparing the logical $\ket{\overline{{++}+}}$ state for the $[[8,3,2]]_1$ code counts as a failure if the output state is a $+1$ eigenstate of the stabilizers in \cref{eq:v1_stabilizers} but not a $+1$ eigenstate of the logical operators $\overline{X}_1, \overline{X}_2$, and $\overline{X}_3$ defined in \cref{eq:v1_logical_x}. 

\Cref{fig:sim_v12_init} shows the logical error rate and acceptance rate for state-preparation circuits of Versions 1 and 2 of the $[[8,3,2]]$ code after $10^6$ Monte Carlo samples. In  these simulations, we apply depolarizing noise with physical error rate $p$ to every CNOT gate and report both the postselected logical error rate 
\begin{equation}\label{eq:logical_error_rate}
    p_{L} = \frac{n_{\mathrm{failure}}}{n_{\mathrm{postselected}}},
\end{equation} 
and the acceptance rate
\begin{equation}\label{eq:acceptance_rate}
    R = \frac{n_{\mathrm{postselected}}}{n_{\mathrm{total}}},
\end{equation}
i.e., the fraction of accepted runs among all runs.

The quantity $p_L$ therefore measures the logical infidelity conditioned on acceptance, whereas $R$ measures the postselection overhead. Because both versions of the $[[8,3,2]]$ code are error-detecting rather than error-correcting, the relevant signature of weak fault tolerance is that accepted runs exhibit $p_L=O(p^2)$. This is what we observe in the left panels of \cref{fig:sim_v12_init}: for $p<0.1$, the postselected logical error rate scales approximately quadratically with the physical error rate, indicating that only weight-$2$ (or higher) faults contribute to accepted-run logical failures in the state-preparation circuits described in \cref{app:B}.

The right panels of \cref{fig:sim_v12_init} show the complementary behavior of the acceptance rate. At low physical error rates, $R$ remains close to $1$ for all preparations, and it decreases monotonically as $p$ increases because single detectable faults are removed by postselection. Thus, the simulations exhibit the expected tradeoff for an error-detecting protocol: postselection suppresses the logical error rate of accepted runs to second order, but it does so at the cost of an acceptance penalty that appears already at first order in $p$. The spread among the acceptance curves is circuit dependent and reflects differences in circuit size and in the number and type of postselection checks. In Version 1, the four acceptance curves remain fairly close, indicating comparable postselection overhead across the four preparations. In Version 2, the spread is more pronounced: the $\ket{\overline{{++}+}}$ preparation has the lowest acceptance, while $\ket{\overline{00+}}$ has the highest acceptance over most of the plotted range. Acceptance and postselected logical fidelity therefore need not track each other monotonically: for example, in Version 2 the $\ket{\overline{000}}$ preparation has a lower postselected logical error rate than $\ket{\overline{00+}}$ over most of the plotted range, despite also having a lower acceptance rate. Lower acceptance therefore does not mean lower fidelity of the accepted states; rather, it means that the circuit detects and discards faulty runs more frequently.

\subsection{Code switching}
\label{sec:sim-code-switch}

\begin{figure*}
    \centering
    \includegraphics[width=0.9\textwidth]{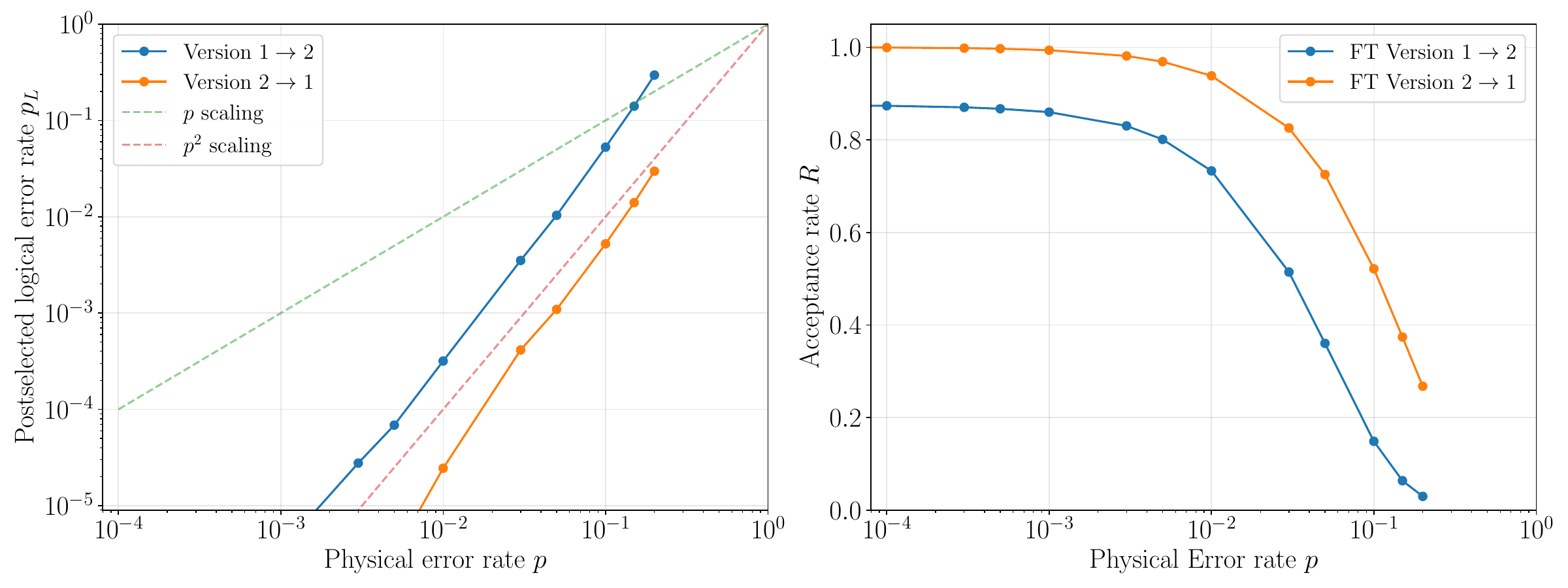}
    \caption{Numerical simulation of the isolated switching-step benchmark. Left: postselected logical error rate for switching in both directions. Right: acceptance rates for the flagged Version~1$\to$2 benchmark and the unflagged Version~2$\to$1 benchmark. We apply depolarizing noise with physical error rate $p$ after each CNOT gate and $q=p/10$ after each single-qubit gate, and assume noise-free feedback and Pauli corrections. In this benchmark, switching from Version~1 to Version~2 uses flagged weight-$4$ Pauli-$Z$ gauge measurements, whereas switching from Version~2 to Version~1 uses weight-$2$ Pauli-$X$ gauge measurements without flag qubits. In the low-noise regime, the postselected logical error rates scale approximately as $p^2$, consistent with weak fault tolerance.}
    \label{fig:sim_switching} 
\end{figure*}

We next benchmark the noisy switching step underlying the code-switching protocol of \cref{sec:4}. Each shot is classified as successful, discarded, or failed. A shot is discarded when the switching procedure detects a fault, e.g., when a flag is triggered during a flagged weight-$4$ gauge measurement or when the output fails the target-code stabilizer or gauge checks. Among the accepted shots, the switching is counted as a failure if the final state no longer matches the target logical state. Thus, the logical error rate $p_L$ characterizes accepted runs, while the acceptance rate $R$ quantifies the postselection overhead.

To isolate the switching step itself, we start from a noiseless encoded source state and inject noise only during switching.
In this benchmark, we do not include the complementary checks of \cref{sec:4}; instead, after the code-switching circuit, we apply noiseless postselection on the target-code stabilizers and fixed gauge operators.
In each direction, we use the representative logical state $\ket{\overline{{++}+}}$, prepared with the corresponding noise-free encoding circuit from \cref{app:B}. We then evaluate the output using the target-code stabilizers, fixed gauge operators, and logical operators. This benchmark therefore tests whether the switching step preserves this logical state under postselection. Because pure logical $X$-type faults act trivially on $\ket{\overline{{++}+}}$, this benchmark should be interpreted as a state-specific diagnostic rather than a complete characterization of the logical channel.

In the benchmarked gauge-measurement step, the two switching directions have different overheads. As shown in \cref{fig:measure}, switching from Version~1 to Version~2 requires flagged measurements of three weight-$4$ Pauli-$Z$ gauge operators, whereas switching from Version~2 to Version~1 uses only weight-$2$ Pauli-$X$ gauge measurements and therefore needs no flag qubits. This asymmetry is reflected in the simulation results.

\Cref{fig:sim_switching} summarizes $10^6$ Monte Carlo samples. We use depolarizing noise with physical error rate $p$ after each CNOT gate and $q=p/10$ after each single-qubit gate, and we again compute the postselected logical error rate $p_L$ and the acceptance rate $R$ using \cref{eq:logical_error_rate,eq:acceptance_rate}. The left panel shows $p_L$ for switching in both directions. In both cases, the postselected logical error rate scales approximately quadratically with $p$ in the low-noise regime, consistent with weak fault tolerance. Moreover, the Version~2$\to$1 curve lies below the Version~1$\to$2 curve across the plotted range, as expected from the simpler unflagged weight-$2$ measurements used in the reverse direction.

The right panel of \cref{fig:sim_switching} shows the acceptance rates for the flagged Version~1$\to$2 protocol and the unflagged Version~2$\to$1 protocol. 
At the lowest plotted physical error rates, both acceptance rates are high, with the Version~2$\to$1 curve closer to unity than the Version~1$\to$2 curve, and both
decrease monotonically as detectable single faults are removed by postselection. This first-order reduction in $R$ is the expected cost of an error-detecting protocol, which discards a fraction of shots already at $O(p)$. The two acceptance curves therefore quantify the postselection overhead of the isolated 
switching-step benchmarks, after immediate postselection on the target-code fixed gauge operators and stabilizers. The Version~2$\to$1 curve lies above the Version~1$\to$2 curve across the plotted range, consistent with the simpler unflagged weight-$2$ measurements used in that direction.
In the full switching protocol of \cref{sec:4}, we additionally include the complementary checks to ensure that the target code is reached in the correct gauge and stabilizer sector before subsequent logical gates are applied.

\section{Grover's Algorithm with code-switching}
\label{sec:Grover}

\begin{figure}[b]
    \centering
    \includegraphics[width=0.45\textwidth]{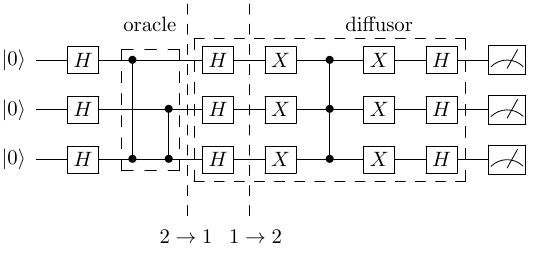}
    \caption{Logical Grover circuit for the two marked strings $011$ and $101$ among eight candidates. The vertical dashed lines indicate the two code-switching locations used in the encoded implementation. The initial $\overline{H}^{\otimes 3}$ layer is absorbed into preparation of $\ket{\overline{{++}+}}$ in Version~2, and the final $\overline{H}^{\otimes 3}$ layer is absorbed into destructive $X$-basis readout.}
    \label{fig:grover_circuit}
\end{figure}

\begin{figure*}[t]
    \centering
    \includegraphics[width=0.9\textwidth]{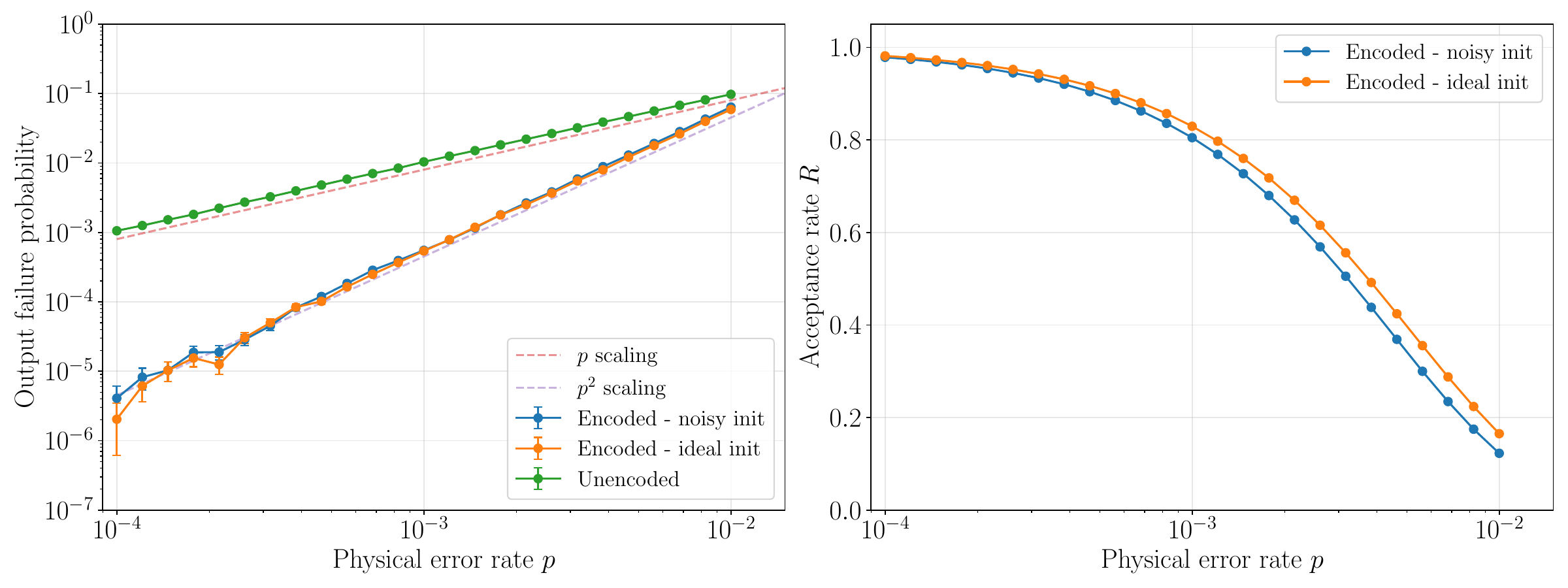}
    \caption{
    Simulation results for Grover's algorithm with $q=p$. Left: output failure probability for the unencoded three-qubit circuit and postselected decoded-output failure probabilities for two encoded implementations: one that assumes the Version~2 input state $\ket{\overline{{++}+}}$ is available, and one that includes its fault-tolerant preparation 
    together with the additional flagged $G^Z_5$ and $G^Z_6$  checks. 
     Right: acceptance rates for the two encoded implementations. In the low-noise regime, the unencoded circuit exhibits approximately linear scaling with $p$, whereas both encoded implementations exhibit approximately quadratic scaling.
    \label{fig:grover_sim}}
\end{figure*}

In this section, we use a three-logical-qubit instance of Grover's algorithm~\cite{grover1996fast,Grover:1997aa} as an algorithmic benchmark for our code-switching scheme. Grover search has also been employed as a small-scale benchmark for computation with error detection using the $[[n,n-2,2]]$ and $[[8,3,2]]$ codes~\cite{Pokharel2024npj,GinsbergPatel2025,Buttetal2026}. 
The algorithm searches for a set of marked strings $M \subseteq \{0,1\}^n$ in an unstructured database of size $N=2^n$, where the oracle operator $O$ is defined by
\beq
    O\ket{x}=\begin{cases}
        -\ket{x}, & x\in M, \\
        \ket{x}, & \text{otherwise},
    \end{cases}
\eeq
for any $n$-bit string $x$. A classical algorithm requires $O(N/|M|)$ oracle queries, whereas Grover's algorithm achieves a quadratic speedup using $O(\sqrt{N/|M|})$ oracle queries~\cite{Boyer:96,Biham:1999ye}.

\Cref{fig:grover_circuit} shows the case $M=\{101,011\}$ for $n=3$. In the ideal noiseless circuit, one Grover iteration maps the initial uniform superposition to $(\ket{101}+\ket{011})/\sqrt{2}$, so no other output bit string occurs upon measurement. In the simulations below, any other output bit string is counted as an output failure.

Grover's algorithm is a useful small-scale demonstration of our protocol because its logical circuit uses both a logical $\overline{H}$ gate and a logical $\overline{\mathrm{CCZ}}$ gate (see \cref{fig:grover_circuit}). Within the fault-tolerant gate sets available in a single version of the $[[8,3,2]]$ code considered here, implementing both requires switching between the two versions. 
To evaluate the algorithm's performance with code switching, we simulated three cases using \texttt{Qiskit} AerSimulator~\cite{QiskitAer}: the unencoded three-qubit circuit, an encoded implementation that assumes the Version~2 input state $\ket{\overline{{++}+}}$ is available, and an encoded implementation that includes the fault-tolerant preparation circuit for $\ket{\overline{{++}+}}$ 
 and $G^Z_6$  checks.
For this benchmark, we use the same depolarizing noise model as in \cref{sec:sim}, except with a different value for $q$: after each two-qubit gate we apply depolarizing noise with physical error rate $p$, and after each single-qubit gate we apply depolarizing noise with rate $q=p$. 
Measurement, feedforward, and gauge corrections are treated as ideal. For each physical error rate between $10^{-4}$ and $10^{-2}$, we sampled $10^6$ shots.

In the encoded simulations, the final logical $\overline{H}^{\otimes 3}$ layer is absorbed into destructive $X$-basis measurements of all data qubits. Let $m_j\in\{-1,+1\}$ denote the $X$-basis outcome on physical qubit $j$,
and first postselect on
\beq
\prod_{j=0}^7 m_j=+1,
\eeq
which is the $X^{\otimes 8}$ stabilizer check. 
For Version~2, the corresponding logical 
$X$-eigenvalue outcomes are
\beq
\begin{aligned}
    b_1 &= m_0  m_1  m_2  m_3\\
    b_2 &= m_0  m_1  m_4  m_5\\
    b_3 &= m_0  m_2  m_4  m_6,
\end{aligned}
\eeq
which measure the Version~2 logical operators $\overline{X}_1$, $\overline{X}_2$, and $\overline{X}_3$. 
Equivalently, the decoded logical bits are
\beq
s_i=\frac{1-b_i}{2}\in\{0,1\}.
\eeq
A shot is counted as successful if the decoded bit string 
$s_1s_2s_3$ is $101$ or $011$. The unencoded benchmark is evaluated analogously after absorbing the final Hadamards into $X$-basis readout. 
Postselection is applied to all flag, complementary check, and stabilizer measurements, as well as to the ancilla checks 
in the weakly fault-tolerant $\widetilde{XX}$ and $\widetilde{ZZ}$ gadgets used to implement the logical Hadamards.

In the encoded case with noisy state preparation, postselection is also applied to the auxiliary measurements immediately after the state-preparation circuit in \cref{fig:v2-+++} together with the additional flagged $G^Z_5$ and $G^Z_6$ checks. 
In the encoded case with noiseless state preparation, no such immediate postselection step is required because no faults are inserted during state preparation. 
We verified by noiseless simulation that the encoded physical circuits, together with this $X$-basis readout and classical postprocessing, reproduce the output of the ideal logical circuit.

Accordingly, the left panel of \cref{fig:grover_sim} reports the output failure probability of the unencoded circuit and the postselected decoded-output failure probabilities of the two encoded implementations. The unencoded circuit exhibits approximately linear scaling with the physical error rate, while both encoded implementations exhibit approximately quadratic scaling in the low-noise regime. The encoded case that includes the noisy initial state preparation circuit accounts for the additional overhead of preparing the Version~2 input state, whereas the case that assumes noiseless initial state preparation of the logical $\ket{\overline{{++}+}}$ state isolates the contribution of the Grover and code-switching steps themselves.

The right panel of \cref{fig:grover_sim} shows the corresponding acceptance rates for the two encoded implementations. In both cases, the acceptance remains close to unity at low physical error rate and then decreases monotonically as detectable faults are removed by postselection. Including the fault-tolerant state-preparation circuit lowers the acceptance rate because it introduces additional noisy locations and additional postselection checks. This first-order acceptance penalty is the expected cost of the postselected subroutines, including state preparation, 
the weakly fault-tolerant $\widetilde{XX}$ and $\widetilde{ZZ}$ gadgets used to implement the logical Hadamards,
and code switching: the accepted-run output failure is suppressed to second order, but only a fraction $R$ of shots are retained.

\section{Conclusion}
In this paper, we developed a fault-tolerant, error-detecting protocol for switching between two versions of the $[[8,3,2]]$ code and used it to assemble a universal logical gate set. Numerical simulations of state preparation and code switching
show quadratic suppression of postselected logical error rates, 
and the three-logical-qubit Grover benchmark shows quadratic suppression of the postselected decoded-output failure probability, consistent with weak fault tolerance.

We showed that both versions arise from the same $[[8,3,3,2]]$ subsystem code, obtained by gauging out three logical qubits of the $[[8,6,2]]$ code after a noncanonical choice of logical basis. Fixing the gauge qubits in $\ket{0_g}$ yields Version 2, namely the $[[8,3,2]]$ code with logical $\overline{\mathrm{CCZ}}$, $\overline{\mathrm{CZ}}$, $\overline{\mathrm{CNOT}}$, and $\overline{\mathrm{SWAP}}$ gates. Fixing them in $\ket{+_g}$ yields Version 1, in which the resulting weight-$2$ representatives of the logical operators enable weakly fault-tolerant implementations of $\overline{H}$, $\overline{S}$, and $\overline{\sqrt{X}^\dagger}$ using the $\widetilde{XX}$ and $\widetilde{ZZ}$ rotation gadgets. We also showed that single inter-block logical $\overline{\mathrm{CNOT}}$ gates are available between same-version blocks and, for mixed-version blocks, only in the direction $[[8,3,2]]_2 \to [[8,3,2]]_1$.

Our construction suggests a broader family viewpoint. The code $[[8,3,2]]_2$ belongs to the $[[2^D,D,2]]$ hypercube family \cite{Kubica_2015, vasmer2022morphing} with $D=3$, while the parent $[[8,6,2]]$ code belongs to the $[[n,n-2,2]]$ family. Our use of a noncanonical logical basis raises the question of how to choose logical bases systematically for $[[2^D,2^D-2,2]]$ codes so that $2^D-2-D$ logical qubits may be gauged out while preserving useful transversal structure. Fixing those gauge qubits in $\ket{0_g}$ recovers the $D$-dimensional hypercube code, while fixing them in $\ket{+_g}$ suggests a companion family generalizing our Version 1 construction. More generally, $D$-dimensional hypercube codes support transversal multi-controlled-phase gates of increasing order \cite{webster2022xp,hangleiter2025fault}, providing depth-1 access to higher levels of the Clifford hierarchy. It would therefore be interesting to characterize the corresponding Version 1 family systematically and to determine its transversal logical gate set.\\

\section*{Acknowledgements}
\addcontentsline{toc}{section}{Acknowledgements}
The authors acknowledge Christopher Gerhard, Anirudh Lanka and Juan Garcia Nila for useful discussions. This material is based upon work supported by, or in part by, the U. S. Army Research Laboratory and the U. S. Army Research Office under contract/grant number W911NF2310255 and by NSF Grant FET-2316713.

\bibliography{references}

@misc{gerhard2024weakly,
      title={Weakly Fault-Tolerant Computation in a Quantum Error-Detecting Code}, 
      author={Christopher Gerhard and Todd A. Brun},
      year={2026},
      eprint={2408.14828},
      archivePrefix={arXiv},
      primaryClass={quant-ph},
      url={https://arxiv.org/abs/2408.14828}, 
}

@article{Litinski2019magicstate,
  doi = {10.22331/q-2019-12-02-205},
  url = {https://doi.org/10.22331/q-2019-12-02-205},
  title = {Magic {S}tate {D}istillation: {N}ot as {C}ostly as {Y}ou {T}hink},
  author = {Litinski, Daniel},
  journal = {{Quantum}},
  issn = {2521-327X},
  publisher = {{Verein zur F{\"{o}}rderung des Open Access Publizierens in den Quantenwissenschaften}},
  volume = {3},
  pages = {205},
  month = dec,
  year = {2019}
}

@misc{campbell2016smallest,
  author       = {Campbell, Earl T.},
  title        = {The Smallest Interesting Colour Code},
  year         = {2016},
  url = {https://earltcampbell.com/2016/09/26/the-smallest-interesting-colour-code/}
}

@article{honciuc2024implementing,
  title = {Implementing fault-tolerant non-Clifford gates using the [[8,3,2]] color code},
  author = {Honciuc Menendez, Daniel and Ray, Annie and Vasmer, Michael},
  journal = {Phys. Rev. A},
  volume = {109},
  issue = {6},
  pages = {062438},
  numpages = {13},
  year = {2024},
  month = {Jun},
  publisher = {American Physical Society},
  doi = {10.1103/PhysRevA.109.062438},
  url = {https://link.aps.org/doi/10.1103/PhysRevA.109.062438}
}

@article{Kribs:2005:180501,
	author = {Kribs, David and Laflamme, Raymond and Poulin, David},
	doi = {10.1103/PhysRevLett.94.180501},
	issue = {18},
	journal = {Phys. Rev. Lett.},
	month = {May},
	numpages = {4},
	pages = {180501},
	publisher = {American Physical Society},
	title = {Unified and Generalized Approach to Quantum Error Correction},
	url = {http://link.aps.org/doi/10.1103/PhysRevLett.94.180501},
	volume = {94},
	year = {2005}}

@article{Kosut:2008lq,
	author = {Kosut, Robert L. and Shabani, Alireza and Lidar, Daniel A.},
	date = {2008/01/16/},
	day = {16},
	id = {10.1103/PhysRevLett.100.020502},
	j1 = {PRL},
	journal = {Physical Review Letters},
	journal1 = {Phys. Rev. Lett.},
	month = {01},
	number = {2},
	pages = {020502--},
	publisher = {American Physical Society},
	title = {Robust Quantum Error Correction via Convex Optimization},
	ty = {JOUR},
	url = {http://link.aps.org/doi/10.1103/PhysRevLett.100.020502},
	volume = {100},
	year = {2008}}

@article{wang2024fault,
    author = {Yang Wang and others },
    title = {Fault-tolerant one-bit addition with the smallest interesting color code},
    journal = {Science Advances},
    volume = {10},
    number = {29},
    pages = {eado9024},
    year = {2024},
    doi = {10.1126/sciadv.ado9024},
    URL = {https://www.science.org/doi/abs/10.1126/sciadv.ado9024},
    eprint = {https://www.science.org/doi/pdf/10.1126/sciadv.ado9024}
}

@article{bluvstein2024logical,
    author={Bluvstein, Dolev and others},
    title={Logical quantum processor based on reconfigurable atom arrays},
    journal={Nature},
    year={2024},
    month={Feb},
    day={01},
    volume={626},
    number={7997},
    pages={58-65},
    issn={1476-4687},
    doi={10.1038/s41586-023-06927-3},
    url={https://doi.org/10.1038/s41586-023-06927-3}
}

@article{kubica2015universal,
  title = {Universal transversal gates with color codes: A simplified approach},
  author = {Kubica, Aleksander and Beverland, Michael E.},
  journal = {Phys. Rev. A},
  volume = {91},
  issue = {3},
  pages = {032330},
  numpages = {12},
  year = {2015},
  month = {Mar},
  publisher = {American Physical Society},
  doi = {10.1103/PhysRevA.91.032330},
  url = {https://link.aps.org/doi/10.1103/PhysRevA.91.032330}
}

@article{butt2024fault,
  title = {Fault-Tolerant Code-Switching Protocols for Near-Term Quantum Processors},
  author = {Butt, Friederike and Heu\ss{}en, Sascha and Rispler, Manuel and M\"uller, Markus},
  journal = {PRX Quantum},
  volume = {5},
  issue = {2},
  pages = {020345},
  numpages = {26},
  year = {2024},
  month = {May},
  publisher = {American Physical Society},
  doi = {10.1103/PRXQuantum.5.020345},
  url = {https://link.aps.org/doi/10.1103/PRXQuantum.5.020345}
}

@article{vasmer2022morphing,
  title = {Morphing Quantum Codes},
  author = {Vasmer, Michael and Kubica, Aleksander},
  journal = {PRX Quantum},
  volume = {3},
  issue = {3},
  pages = {030319},
  numpages = {24},
  year = {2022},
  month = {Aug},
  publisher = {American Physical Society},
  doi = {10.1103/PRXQuantum.3.030319},
  url = {https://link.aps.org/doi/10.1103/PRXQuantum.3.030319}
}

@article{chao2018quantum,
  title = {Quantum Error Correction with Only Two Extra Qubits},
  author = {Chao, Rui and Reichardt, Ben W.},
  journal = {Phys. Rev. Lett.},
  volume = {121},
  issue = {5},
  pages = {050502},
  numpages = {5},
  year = {2018},
  month = {Aug},
  publisher = {American Physical Society},
  doi = {10.1103/PhysRevLett.121.050502},
  url = {https://link.aps.org/doi/10.1103/PhysRevLett.121.050502}
}

@article{eastin2009restrictions,
  title = {Restrictions on Transversal Encoded Quantum Gate Sets},
  author = {Eastin, Bryan and Knill, Emanuel},
  journal = {Phys. Rev. Lett.},
  volume = {102},
  issue = {11},
  pages = {110502},
  numpages = {4},
  year = {2009},
  month = {Mar},
  publisher = {American Physical Society},
  doi = {10.1103/PhysRevLett.102.110502},
  url = {https://link.aps.org/doi/10.1103/PhysRevLett.102.110502}
}

@article{kitaev1997quantum,
    doi = {10.1070/RM1997v052n06ABEH002155},
    url = {https://doi.org/10.1070/RM1997v052n06ABEH002155},
    year = {1997},
    month = {dec},
    publisher = {},
    volume = {52},
    number = {6},
    pages = {1191},
    author = {A Yu Kitaev},
    title = {Quantum computations: algorithms and error correction},
    journal = {Russian Mathematical Surveys}
}

@article{aharonov1997fault,
    author = {Aharonov, Dorit and Ben-Or, Michael},
    title = {Fault-Tolerant Quantum Computation with Constant Error Rate},
    journal = {SIAM Journal on Computing},
    volume = {38},
    number = {4},
    pages = {1207-1282},
    year = {2008},
    doi = {10.1137/S0097539799359385},
    URL = {https://doi.org/10.1137/S0097539799359385},
    eprint = {https://doi.org/10.1137/S0097539799359385}
}

@phdthesis{Gottesman:97b,
	address = {Pasadena, CA},
	author = {{D. Gottesman}},
	eprint = {quant-ph/9705052},
	school = {California Institute of Technology},
	title = {{Stabilizer Codes and Quantum Error Correction}},
	year = {1997}}

@article{quan2018fault,
    doi = {10.1088/1751-8121/aaad13},
    url = {https://doi.org/10.1088/1751-8121/aaad13},
    year = {2018},
    month = {feb},
    publisher = {IOP Publishing},
    volume = {51},
    number = {11},
    pages = {115305},
    author = {Quan, Dong-Xiao and Zhu, Li-Li and Pei, Chang-Xing and Sanders, Barry C},
    title = {Fault-tolerant conversion between adjacent Reed–Muller quantum codes based on gauge fixing},
    journal = {Journal of Physics A: Mathematical and Theoretical},
}

@article{bombin2015gauge,
    doi = {10.1088/1367-2630/17/8/083002},
    url = {https://doi.org/10.1088/1367-2630/17/8/083002},
    year = {2015},
    month = {aug},
    publisher = {IOP Publishing},
    volume = {17},
    number = {8},
    pages = {083002},
    author = {Bombín, Héctor},
    title = {Gauge color codes: optimal transversal gates and gauge fixing in topological stabilizer codes},
    journal = {New Journal of Physics}
}

@article{hangleiter2025fault,
  title = {Fault-Tolerant Compiling of Classically Hard Instantaneous Quantum Polynomial Circuits on Hypercubes},
  author = {Hangleiter, Dominik and Kalinowski, Marcin and Bluvstein, Dolev and Cain, Madelyn and Maskara, Nishad and Gao, Xun and Kubica, Aleksander and Lukin, Mikhail D. and Gullans, Michael J.},
  journal = {PRX Quantum},
  volume = {6},
  issue = {2},
  pages = {020338},
  numpages = {50},
  year = {2025},
  month = {May},
  publisher = {American Physical Society},
  doi = {10.1103/PRXQuantum.6.020338},
  url = {https://link.aps.org/doi/10.1103/PRXQuantum.6.020338}
}

@Inbook{greenberger1989going,
    author="Greenberger, Daniel M. and Horne, Michael A. and Zeilinger, Anton",
    editor="Kafatos, Menas",
    title="Going Beyond Bell's Theorem",
    bookTitle="Bell's Theorem, Quantum Theory and Conceptions of the Universe",
    year="1989",
    publisher="Springer Netherlands",
    address="Dordrecht",
    pages="69--72",
    isbn="978-94-017-0849-4",
    doi="10.1007/978-94-017-0849-4_10",
    url="https://doi.org/10.1007/978-94-017-0849-4_10"
}

@article{Grover:1997aa,
	author = {Grover, Lov K.},
	date = {1997/12/08/},
	day = {08},
	doi = {10.1103/PhysRevLett.79.4709},
	id = {10.1103/PhysRevLett.79.4709},
	j1 = {PRL},
	journal = {Physical Review Letters},
	journal1 = {Phys. Rev. Lett.},
	month = {12},
	number = {23},
	pages = {4709--4712},
	publisher = {American Physical Society},
	title = {Quantum Computers Can Search Arbitrarily Large Databases by a Single Query},
	url = {https://link.aps.org/doi/10.1103/PhysRevLett.79.4709},
	volume = {79},
	year = {1997}}

@inproceedings{grover1996fast,
	acmid = {237866},
	address = {New York, NY, USA},
	author = {Grover, Lov K.},
	booktitle = {Proceedings of the Twenty-eighth Annual ACM Symposium on Theory of Computing},
	doi = {10.1145/237814.237866},
	isbn = {0-89791-785-5},
	location = {Philadelphia, Pennsylvania, USA},
	numpages = {8},
	pages = {212--219},
	publisher = {ACM},
	series = {STOC '96},
	title = {A Fast Quantum Mechanical Algorithm for Database Search},
	url = {http://doi.acm.org/10.1145/237814.237866},
	year = {1996}}

@article{shor_scheme_1995,
	author = {Shor, Peter W.},
	doi = {10.1103/PhysRevA.52.R2493},
	journal = pra,
	month = oct,
	number = {4},
	pages = {R2493--R2496},
	title = {Scheme for reducing decoherence in quantum computer memory},
	urldate = {2012-10-12},
	volume = {52},
	year = {1995}}

@article{Steane:96a,
	author = {Steane, A. M.},
	date = {1996/07/29/},
	day = {29},
	id = {10.1103/PhysRevLett.77.793},
	j1 = {PRL},
	journal = {Phys. Rev. Lett.},
	journal1 = {Phys. Rev. Lett.},
	month = {07},
	number = {5},
	pages = {793--797},
	publisher = {American Physical Society},
	title = {Error Correcting Codes in Quantum Theory},
	ty = {JOUR},
	url = {http://link.aps.org/doi/10.1103/PhysRevLett.77.793},
	volume = {77},
	year = {1996}}

@article{Calderbank:1997aa,
	author = {Calderbank, A. R. and Rains, E. M. and Shor, P. W. and Sloane, N. J. A.},
	date = {1997/01/20/},
	day = {20},
	doi = {10.1103/PhysRevLett.78.405},
	id = {10.1103/PhysRevLett.78.405},
	j1 = {PRL},
	journal = {Physical Review Letters},
	journal1 = {Phys. Rev. Lett.},
	month = {01},
	number = {3},
	pages = {405--408},
	publisher = {American Physical Society},
	title = {Quantum Error Correction and Orthogonal Geometry},
	url = {https://link.aps.org/doi/10.1103/PhysRevLett.78.405},
	volume = {78},
	year = {1997}}

@book{Gaitan:book,
	address = {Boca Raton},
	author = {F. Gaitan},
	publisher = {Taylor \& Francis Group},
	title = {Quantum Error Correction and Fault Tolerant Quantum Computing},
	url = {http://books.google.com/books?id=zwvlqspyOK8C},
	year = {2008}}

@book{Lidar-Brun:book,
	address = {{Cambridge, UK}},
	editor = {D.A. Lidar and T.A. Brun},
	publisher = {Cambridge University Press},
	title = {Quantum Error Correction},
	url = {http://www.cambridge.org/9780521897877},
	year = {2013}}

@article{Knill:1997kx,
	author = {Knill, Emanuel and Laflamme, Raymond},
	date = {1997/02/01/},
	day = {01},
	id = {10.1103/PhysRevA.55.900},
	j1 = {PRA},
	journal = {Phys. Rev. A},
	journal1 = {Phys. Rev. A},
	month = {02},
	number = {2},
	pages = {900--911},
	publisher = {American Physical Society},
	title = {Theory of quantum error-correcting codes},
	ty = {JOUR},
	url = {http://link.aps.org/doi/10.1103/PhysRevA.55.900},
	volume = {55},
	year = {1997}}

@inbook{reichardt_fault-tolerance_2005,
	address = {Berlin, Heidelberg},
	author = {Reichardt, Ben W.},
	booktitle = {Automata, Languages and Programming: 33rd International Colloquium, ICALP 2006, Venice, Italy, July 10-14, 2006, Proceedings, Part I},
	da = {2006//},
	editor = {Bugliesi, Michele and Preneel, Bart and Sassone, Vladimiro and Wegener, Ingo},
	id = {Reichardt2006},
	isbn = {978-3-540-35905-0},
	pages = {50--61},
	publisher = {Springer Berlin Heidelberg},
	title = {Fault-Tolerance Threshold for a Distance-Three Quantum Code},
	ty = {CHAP},
	url = {http://dx.doi.org/10.1007/11786986_6},
	year = {2006}}

@article{Knill:2000dq,
	author = {Knill, Emanuel and Laflamme, Raymond and Viola, Lorenza},
	date = {2000/03/13/},
	day = {13},
	id = {10.1103/PhysRevLett.84.2525},
	j1 = {PRL},
	journal = {{Phys.~Rev.~Lett.}},
	journal1 = {Phys. Rev. Lett.},
	month = {03},
	number = {11},
	pages = {2525--2528},
	publisher = {American Physical Society},
	title = {Theory of Quantum Error Correction for General Noise},
	ty = {JOUR},
	url = {http://link.aps.org/doi/10.1103/PhysRevLett.84.2525},
	volume = {84},
	year = {2000}}

@article{Bacon:05,
	author = {Bacon, Dave},
	doi = {10.1103/PhysRevA.73.012340},
	issue = {1},
	journal = {Phys. Rev. A},
	month = {Jan},
	numpages = {13},
	pages = {012340},
	publisher = {American Physical Society},
	title = {Operator quantum error-correcting subsystems for self-correcting quantum memories},
	url = {https://link.aps.org/doi/10.1103/PhysRevA.73.012340},
	volume = {73},
	year = {2006}}

@article{poulin_stabilizer_2005,
	author = {Poulin, David},
	doi = {10.1103/PhysRevLett.95.230504},
	issue = {23},
	journal = {Phys. Rev. Lett.},
	month = {Dec},
	numpages = {4},
	pages = {230504},
	publisher = {American Physical Society},
	title = {Stabilizer Formalism for Operator Quantum Error Correction},
	url = {http://link.aps.org/doi/10.1103/PhysRevLett.95.230504},
	volume = {95},
	year = {2005}}

@article{Gottesman:1996fk,
	author = {Gottesman, D.},
	date = {1996/09/01/},
	day = {01},
	doi = {10.1103/PhysRevA.54.1862},
	j1 = {PRA},
	journal = {{Phys. Rev. A}},
	journal1 = {Phys. Rev. A},
	month = {09},
	number = {3},
	pages = {1862},
	publisher = {American Physical Society},
	title = {Class of quantum error-correcting codes saturating the quantum Hamming bound},
	ty = {JOUR},
	volume = {54},
	year = {1996}}

@article{Aliferis:05,
	author = {P. Aliferis and D. Gottesman and J. Preskill},
	journal = {{Quant. Inf. Comput.}},
	pages = {97},
	title = {Quantum accuracy threshold for concatenated distance-3 codes},
	url = {https://arxiv.org/abs/quant-ph/0504218},
	volume = {6},
	year = {2006}}

@article{Knill:05,
	author = {Knill, E.},
	date = {2005/03/03/print},
	day = {03},
	isbn = {0028-0836},
	journal = {Nature},
	l3 = {http://www.nature.com/nature/journal/v434/n7029/suppinfo/nature03350_S1.html},
	m3 = {10.1038/nature03350},
	month = {03},
	number = {7029},
	pages = {39--44},
	title = {Quantum computing with realistically noisy devices},
	ty = {JOUR},
	url = {http://dx.doi.org/10.1038/nature03350},
	volume = {434},
	year = {2005}}

@article{Campbell:2017aa,
	author = {Campbell, Earl T. and Terhal, Barbara M. and Vuillot, Christophe},
	date = {2017/09/13/online},
	day = {13},
	journal = {Nature},
	l3 = {10.1038/nature23460; https://www.nature.com/articles/nature23460#supplementary-information},
	month = {09},
	pages = {172 EP -},
	publisher = {Macmillan Publishers Limited, part of Springer Nature. All rights reserved. SN -},
	title = {Roads towards fault-tolerant universal quantum computation},
	ty = {JOUR},
	url = {http://dx.doi.org/10.1038/nature23460},
	volume = {549},
	year = {2017}}

@article{Shor:96,
	author = {P. W. Shor},
	isbn = {0272-5428},
	journal = {Proceedings of 37th Conference on Foundations of Computer Science},
	pages = {56--65},
	title = {Fault-tolerant quantum computation},
	ty = {CONF},
	url = {http://ieeexplore.ieee.org/document/548464/},
	year = {1996},
	year1 = {14-16 Oct 1996}}

@inproceedings{Boykin:99,
	address = {Los Alamitos, CA},
	author = {{P. Boykin, T. Mor, M. Pulver, V. Roychowdhury, and F. Vatan}},
	booktitle = {{40th Annual Symposium on Foundations of Computer Science}},
	eprint = {quant-ph/9906054},
	pages = {486},
	publisher = {{IEEE Comput. Soc.}},
	title = {{On universal and fault-tolerant quantum computing: a novel basis and a new constructive proof of universality for Shor's basis}},
	year = {1999}}

@article{Bravyi:2005aa,
	author = {Bravyi, Sergey and Kitaev, Alexei},
	date = {2005/02/22/},
	day = {22},
	doi = {10.1103/PhysRevA.71.022316},
	id = {10.1103/PhysRevA.71.022316},
	j1 = {PRA},
	journal = {Physical Review A},
	journal1 = {Phys. Rev. A},
	month = {02},
	number = {2},
	pages = {022316--},
	publisher = {American Physical Society},
	title = {Universal quantum computation with ideal Clifford gates and noisy ancillas},
	url = {https://link.aps.org/doi/10.1103/PhysRevA.71.022316},
	volume = {71},
	year = {2005}}

@article{Anderson:2014aa,
	author = {Anderson, Jonas T. and Duclos-Cianci, Guillaume and Poulin, David},
	date = {2014/08/20/},
	day = {20},
	doi = {10.1103/PhysRevLett.113.080501},
	id = {10.1103/PhysRevLett.113.080501},
	j1 = {PRL},
	journal = {Physical Review Letters},
	journal1 = {Phys. Rev. Lett.},
	month = {08},
	number = {8},
	pages = {080501--},
	publisher = {American Physical Society},
	title = {Fault-Tolerant Conversion between the Steane and Reed-Muller Quantum Codes},
	url = {https://link.aps.org/doi/10.1103/PhysRevLett.113.080501},
	volume = {113},
	year = {2014}}

@misc{vezvaee2025surfacecodescalingheavyhex,
	archiveprefix = {arXiv},
	author = {Arian Vezvaee and Cesar Benito and Mario Morford-Oberst and Alejandro Bermudez and Daniel A. Lidar},
	eprint = {2510.18847},
	primaryclass = {quant-ph},
	title = {Surface code scaling on heavy-hex superconducting quantum processors},
	url = {https://arxiv.org/abs/2510.18847},
	year = {2025}}

@article{Acharya:2025aa,
	author = {{R. Acharya \textit{et al.}}},
	date = {2025/02/01},
	doi = {10.1038/s41586-024-08449-y},
	id = {Acharya2025},
	isbn = {1476-4687},
	journal = {Nature},
	number = {8052},
	pages = {920--926},
	title = {Quantum error correction below the surface code threshold},
	url = {https://doi.org/10.1038/s41586-024-08449-y},
	volume = {638},
	year = {2025}}

@article{Lacroix:2025aa,
	author = {{N. Lacroix \textit{et al.}}},
	date = {2025/09/01},
	doi = {10.1038/s41586-025-09061-4},
	id = {Lacroix2025},
	isbn = {1476-4687},
	journal = {Nature},
	number = {8081},
	pages = {614--619},
	title = {Scaling and logic in the colour code on a superconducting quantum processor},
	url = {https://doi.org/10.1038/s41586-025-09061-4},
	volume = {645},
	year = {2025}}

@article{Eickbusch:2025aa,
	author = {{A. Eickbusch \textit{et al.}}},
	date = {2025/12/01},
	doi = {10.1038/s41567-025-03070-w},
	id = {Eickbusch2025},
	isbn = {1745-2481},
	journal = {Nature Physics},
	number = {12},
	pages = {1994--2001},
	title = {Demonstration of dynamic surface codes},
	url = {https://doi.org/10.1038/s41567-025-03070-w},
	volume = {21},
	year = {2025}}

@article{Bluvstein:2026aa,
	author = {{D. Bluvstein \textit{et al.}}},
	date = {2026/01/01},
	doi = {10.1038/s41586-025-09848-5},
	id = {Bluvstein2026},
	isbn = {1476-4687},
	journal = {Nature},
	number = {8095},
	pages = {39--46},
	title = {A fault-tolerant neutral-atom architecture for universal quantum computation},
	url = {https://doi.org/10.1038/s41586-025-09848-5},
	volume = {649},
	year = {2026}}

@article{DiVincenzo:96,
	author = {DiVincenzo, David P. and Shor , Peter W.},
	date = {1996/10/07/},
	day = {07},
	doi = {10.1103/PhysRevLett.77.3260},
	id = {10.1103/PhysRevLett.77.3260},
	j1 = {PRL},
	journal = {Physical Review Letters},
	journal1 = {Phys. Rev. Lett.},
	month = {10},
	number = {15},
	pages = {3260--3263},
	publisher = {American Physical Society},
	title = {Fault-Tolerant Error Correction with Efficient Quantum Codes},
	url = {https://link.aps.org/doi/10.1103/PhysRevLett.77.3260},
	volume = {77},
	year = {1996}}

@article{Singkanipa2025familiesofdd,
	author = {Singkanipa, Phattharaporn and Xia, Zihan and Lidar, Daniel A.},
	doi = {10.22331/q-2025-08-05-1821},
	issn = {2521-327X},
	journal = {{Quantum}},
	month = aug,
	pages = {1821},
	publisher = {{Verein zur F{\"{o}}rderung des Open Access Publizierens in den Quantenwissenschaften}},
	title = {Families of {$d=2$} 2{D} subsystem stabilizer codes for universal {H}amiltonian quantum computation with two-body interactions},
	url = {https://doi.org/10.22331/q-2025-08-05-1821},
	volume = {9},
	year = {2025}}

@article{Tan2024compilingquantum,
  author = {Tan, Daniel Bochen and Bluvstein, Dolev and Lukin, Mikhail D. and Cong, Jason},
  title = {Compiling Quantum Circuits for Dynamically Field-Programmable Neutral Atoms Array Processors},
  journal = {Quantum},
  volume = {8},
  pages = {1281},
  year = {2024},
  doi = {10.22331/q-2024-03-14-1281}
}

@article{Bluvstein2022coherent,
  author = {Bluvstein, Dolev and Levine, Harry and Semeghini, Giulia and Wang, Tout T. and Ebadi, Sepehr and Kalinowski, Marcin and Keesling, Alexander and Maskara, Nishad and Pichler, Hannes and Greiner, Markus and Vuleti{\'c}, Vladan and Lukin, Mikhail D.},
  title = {A quantum processor based on coherent transport of entangled atom arrays},
  journal = {Nature},
  volume = {604},
  pages = {451--456},
  year = {2022},
  doi = {10.1038/s41586-022-04592-6}
}

@article{Pino2021qccd,
  author = {Pino, J. M. and Dreiling, J. M. and Figgatt, C. and Gaebler, J. P. and Moses, S. A. and Allman, M. S. and Baldwin, C. H. and Foss-Feig, M. and Hayes, D. and Mayer, K. and Ryan-Anderson, C. and Neyenhuis, B.},
  title = {Demonstration of the trapped-ion quantum CCD computer architecture},
  journal = {Nature},
  volume = {592},
  pages = {209--213},
  year = {2021},
  doi = {10.1038/s41586-021-03318-4}
}

@inproceedings{Li2019sabre,
  author = {Li, Gushu and Ding, Yufei and Xie, Yuan},
  title = {Tackling the Qubit Mapping Problem for NISQ-Era Quantum Devices},
  booktitle = {Proceedings of the Twenty-Fourth International Conference on Architectural Support for Programming Languages and Operating Systems},
  pages = {1001--1014},
  year = {2019},
  doi = {10.1145/3297858.3304023}
}

@inproceedings{Cowtan2019routing,
  author = {Cowtan, Alexander and Dilkes, Silas and Duncan, Ross and Krajenbrink, Alexandre and Simmons, Will and Sivarajah, Seyon},
  title = {On the Qubit Routing Problem},
  booktitle = {14th Conference on the Theory of Quantum Computation, Communication and Cryptography (TQC 2019)},
  series = {Leibniz International Proceedings in Informatics (LIPIcs)},
  volume = {135},
  pages = {5:1--5:32},
  year = {2019},
  doi = {10.4230/LIPIcs.TQC.2019.5}
}

@misc{PECOSPackage,
  author       = {Ryan-Anderson, Ciar\'{a}n},
  title        = {{PECOS}: Performance Estimator of Codes On Surfaces},
  year         = {2019},
  publisher    = {GitHub},
  journal      = {GitHub repository},
  howpublished = {\url{https://github.com/PECOS-packages/PECOS}}
}

@article{Pokharel2024npj,
	author = {Pokharel, Bibek and Lidar, Daniel},
	journal = {npj Quantum Information},
	month = {02},
	title = {{Better-than-classical Grover search via quantum error detection and suppression}},
	url = {https://www.nature.com/articles/s41534-023-00794-6},
	volume = {10},
	year = {2024}}

@article{Buttetal2026,
  author  = {Butt, Friederike and Pogorelov, Ivan and Freund, Robert and Steiner, Alex and Meyer, Marcel and Monz, Thomas and M\"{u}ller, Markus},
  title   = {Demonstration of measurement-free universal logical quantum computation},
  journal = {Nature Communications},
  volume  = {17},
  pages   = {995},
  year    = {2026},
  url     = {https://doi.org/10.1038/s41467-026-68533-x}
}

@article{GinsbergPatel2025,
  author  = {Ginsberg, Tom and Patel, Viren},
  title   = {Quantum Error Detection For Early Term Fault-Tolerant Quantum Algorithms},
  journal = {arXiv preprint arXiv:2503.10790},
  year    = {2025},
  url     = {https://arxiv.org/abs/2503.10790}
}

@article{Boyer:96,
	author = {Boyer, Michel and Brassard, Gilles and Hoyer, Peter and Tapp, Alain},
	isbn = {1521-3978},
	journal = {Fortschritte der Physik},
	journal1 = {Fortschr. Phys.},
	number = {4-5},
	pages = {493--505},
	publisher = {WILEY-VCH Verlag Berlin GmbH},
	title = {Tight Bounds on Quantum Searching},
	ty = {JOUR},
	url = {http://dx.doi.org/10.1002/(SICI)1521-3978(199806)46:4/5<493::AID-PROP493>3.0.CO;2-P},
	volume = {46},
	year = {1998},
	year1 = {1998/06/01}}

@article{Biham:1999ye,
	author = {Biham, Eli and Biham, Ofer and Biron, David and Grassl, Markus and Lidar, Daniel A.},
	date = {1999/10/01/},
	day = {01},
	id = {10.1103/PhysRevA.60.2742},
	j1 = {PRA},
	journal = {Physical Review A},
	journal1 = {Phys. Rev. A},
	month = {10},
	number = {4},
	pages = {2742--2745},
	publisher = {American Physical Society},
	title = {Grover's quantum search algorithm for an arbitrary initial amplitude distribution},
	ty = {JOUR},
	url = {http://link.aps.org/doi/10.1103/PhysRevA.60.2742},
	volume = {60},
	year = {1999}}

@misc{QiskitAer,
  author       = {{Qiskit contributors}},
  title        = {Qiskit Aer: A high performance simulator for quantum circuits},
  year         = {2024},
  publisher    = {GitHub},
  journal      = {GitHub repository},
  url          = {https://github.com/Qiskit/qiskit-aer}
}

@article{webster2022xp,
  title={The XP stabiliser formalism: a generalisation of the pauli stabiliser formalism with arbitrary phases},
  author={Webster, Mark A and Brown, Benjamin J and Bartlett, Stephen D},
  journal={Quantum},
  volume={6},
  pages={815},
  year={2022},
  publisher={Verein zur F{\"o}rderung des Open Access Publizierens in den Quantenwissenschaften},
  url={https://doi.org/10.22331/q-2022-09-22-815}
}

@article{Kubica_2015,
doi = {10.1088/1367-2630/17/8/083026},
url = {https://doi.org/10.1088/1367-2630/17/8/083026},
year = {2015},
month = {aug},
publisher = {IOP Publishing},
volume = {17},
number = {8},
pages = {083026},
author = {Kubica, Aleksander and Yoshida, Beni and Pastawski, Fernando},
title = {Unfolding the color code},
journal = {New Journal of Physics},
}

\appendix
\setcounter{equation}{0}

\section{Fault-tolerant implementations of two-qubit rotation gates and Hadamard gate}
\label{app:A}

\begin{figure}[ht]
\centering
    \includegraphics[width=\linewidth]{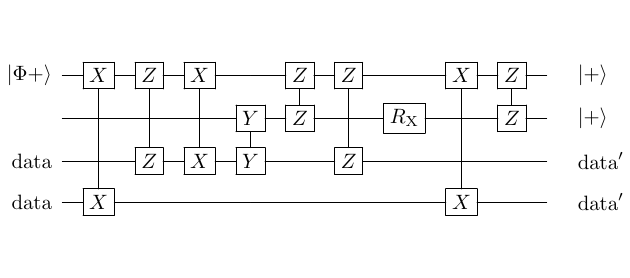}
    \includegraphics[width=\linewidth]{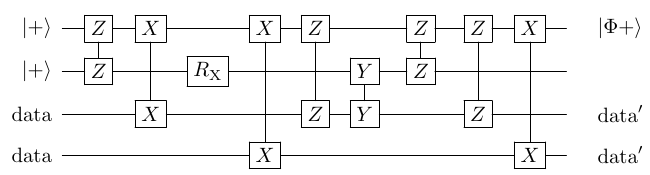}
    \includegraphics[width=\linewidth]{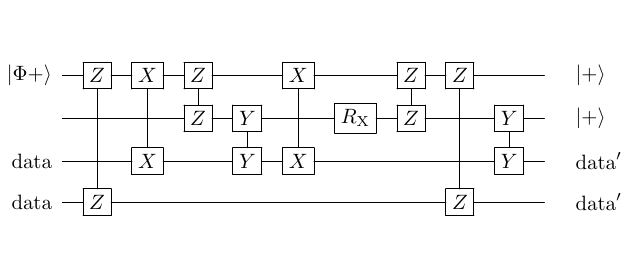}
    \includegraphics[width=\linewidth]{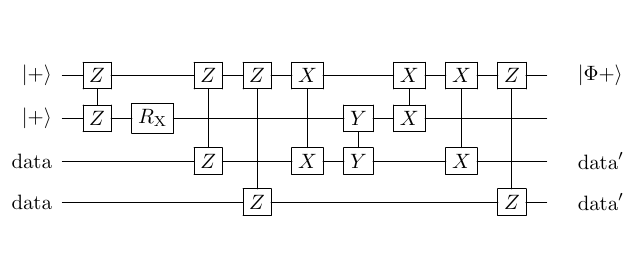}
\caption{
Weakly fault-tolerant gadgets for the $XX$ (top two panels) and $ZZ$ rotations (bottom two panels). The top and third panels consume a Bell pair and return the ancillas in $\ket{++}$; the second and bottom panels consume $\ket{++}$ and return a Bell pair. Alternating the two versions allows successive $XX$ and $ZZ$ gadgets to reuse ancillas without an explicit re-preparation step.}
\label{fig:FTXXZZ}
\end{figure}

\Cref{fig:FTXXZZ} is the fault-tolerant implementation of the $\widetilde{XX}$ and $\widetilde{ZZ}$ gates taken from Ref.~\cite{gerhard2024weakly}. Both implementations have two versions. In the first version, ancillas are prepared in a Bell state and end in a $\ket{++}$ state, while in the second version, ancillas are prepared in the $\ket{++}$ state and end in a Bell state. Hence, alternating between two versions avoids the need to reprepare the ancillas. After each circuit, Pauli operations are applied according to Table \ref{tab:pauli-rec}.

\begin{table}[h]
\centering
\begin{tabular}{|c|c|c|}
\hline
 & $\widetilde{ZZ}$ rotation & $\widetilde{XX}$ rotation \\
\hline
$|\Phi_+\rangle$ & $ZIZI$ & $IZYI$ \\
\hline
$|++\rangle$ & $XIXI$ & $YIYI$ \\
\hline
\end{tabular}
\caption{Pauli recovery operations to apply after each weakly fault-tolerant $\widetilde{XX}$ or $\widetilde{ZZ}$ gate.
The row is determined by the state the ancilla qubits are indirectly in before the gate,
and the column by which rotation gate we are implementing.
}
\label{tab:pauli-rec}
\end{table}

The logical Hadamard gate can be implemented by using the decomposition in \cref{eq:hadamard}. To avoid re-preparing the ancillas between successive two-qubit-rotation gadgets, one may choose the gadget versions so that the ancilla output of one gadget is the required ancilla input of the next: use a Bell-to-$\ket{++}$ version of $\widetilde{ZZ}$, followed by a $\ket{++}$-to-Bell version of $\widetilde{XX}$, followed by a Bell-to-$\ket{++}$ version of $\widetilde{ZZ}$.

To characterize the encoded Hadamard gate, we performed Monte Carlo sampling with two-qubit depolarizing noise applied to every physical two-qubit gate in the circuit; measurements were assumed ideal. For each sampled fault pattern, the induced Pauli error is propagated through the Clifford circuit, the run is postselected on the trivial stabilizer syndrome and on the ancillas returning to the expected final state, and the resulting logical action is compared with the ideal encoded Hadamard on the chosen logical qubit. One could also choose the opposite versions of the $\widetilde{XX}$ and $\widetilde{ZZ}$ gadgets, but then the final ancillas would need to be checked in the Bell basis rather than in the $X$ basis.

\begin{figure*}
    \centering
    \includegraphics[width=\textwidth]{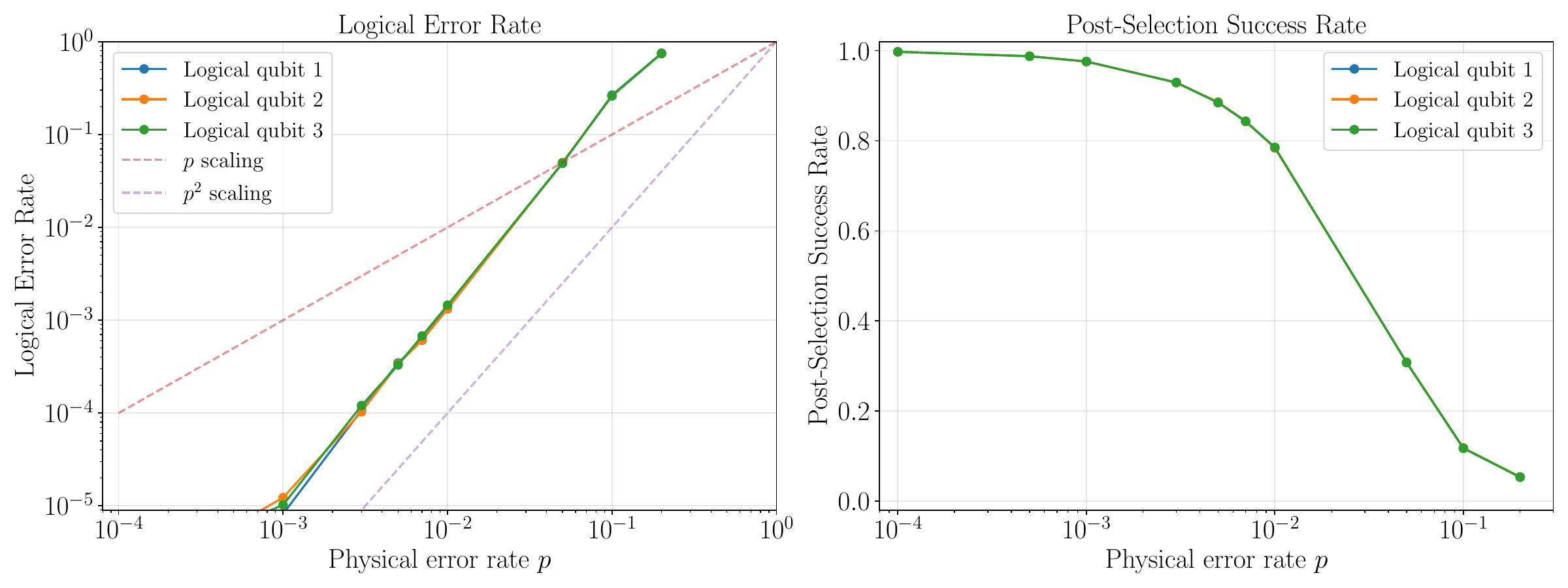}
    \caption{
    Simulation results for the encoded Hadamard gate on the three logical qubits. In the low-noise regime, the postselected logical error rate is approximately quadratic in the physical two-qubit error rate, consistent with weak fault tolerance. The right panel shows the corresponding postselection success rate.}
    \label{fig:sim_hadamard}
\end{figure*}

The resulting postselected logical error rates and postselection success rates for the three logical qubits are shown in \cref{fig:sim_hadamard}. In the low-noise regime, all three implementations exhibit approximately quadratic suppression of the logical error rate with the physical two-qubit error rate, consistent with weak fault tolerance.

\section{Fault-tolerant encoding circuits for both versions of the $[[8,3,2]]$ code}
\label{app:B}

In this appendix, we collect the explicit fault-tolerant state-preparation circuits used throughout the paper. We consider the logical basis states $\ket{\overline{000}}$ and $\ket{\overline{{++}+}}$ in both versions of the $[[8,3,2]]$ code, as well as mixed logical-basis states. For the latter, we use the shorthand $\ket{\overline{+^20^1}}$ and $\ket{\overline{+^10^2}}$ to indicate only the numbers of logical $\ket{\overline{+}}$ and $\ket{\overline{0}}$ states, not their ordering. Without additional gauge fixing, Version 1 naturally supports states of the form $\ket{\overline{+^20^1}}$, while Version 2 naturally supports states of the form $\ket{\overline{+^10^2}}$. The opposite pairing is obtained more efficiently by preparing a partially gauge-fixed intermediate state and then performing only the final gauge-fixing step, rather than a full code switch.

\subsection{GHZ and dual-GHZ resource states}

Most of the encoding circuits below reduce, after Gaussian elimination of the stabilizer generators, to standard GHZ states, dual GHZ states, Bell pairs, or tensor products thereof. An $n$-qubit GHZ state \cite{greenberger1989going} in the computational basis is
\beq
\lvert \mathrm{GHZ}_n \rangle
= \frac{1}{\sqrt{2}}
\left(
  \lvert 0 \rangle^{\otimes n}
  +
  \lvert 1 \rangle^{\otimes n}
\right),
\eeq
with $+1$ eigenoperators $X^{\otimes n}$ and $Z_0 Z_j$ for $j=1,\dots,n-1$. The corresponding dual GHZ state in the $X$ basis is
\beq
\lvert \mathrm{Dual\ GHZ}_n \rangle
= \frac{1}{\sqrt{2}}
\left(
  \lvert + \rangle^{\otimes n}
  +
  \lvert - \rangle^{\otimes n}
\right),
\eeq
with $+1$ eigenoperators $Z^{\otimes n}$ and $X_0 X_j$ for $j=1,\dots,n-1$.

\begin{figure}
    \centering
    \includegraphics[width=0.5\linewidth]{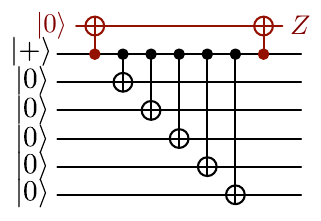}
    \caption{Fault-tolerant GHZ state encoding circuit \cite{chao2018quantum}. The black part is the encoding circuit, and the red part checks for error propagation.}
    \label{fig:ft_ghz}
\end{figure}

The fault-tolerant preparation circuit for $\ket{\mathrm{GHZ}_n}$ is shown in \cref{fig:ft_ghz}. The dual GHZ state is prepared by the same construction after interchanging the $X$ and $Z$ bases, i.e., by replacing $\ket{0}\leftrightarrow\ket{+}$, reversing the CNOT directions, and measuring in the $X$ basis instead of the $Z$ basis.

\begin{figure*}[ht]
    \centering
    \subfigure[$\ket{\overline{000}}$ \label{fig:v1-000}]{
        \resizebox{0.28\linewidth}{!}{
$        \Qcircuit @C=0.9em @R=1.1em{
& &\lstick{\ket{0}} &\qw &\targ &\qw&\qw& \qw & \targ&\qw&Z\\
&\lstick{\ket{+}} &\qw & \qw& \ctrl{-1}&\ctrl{4}& \ctrl{1}& \ctrl{2} & \ctrl{-1}&\qw   \\
&\lstick{\ket{0}} &\qw & \qw& \qw & \qw& \targ& \qw&\qw&\qw\\
&\lstick{\ket{0}} &\qw & \qw& \qw   & \qw& \qw & \targ& \qw&\qw\\
&\lstick{\ket{+}} &\qw & \qw& \ctrl{4}  &\qw & \qw& \qw& \qw&\qw\\ 
&\lstick{\ket{0}} &\qw & \qw&\qw& \targ& \qw & \qw& \qw&\qw\\
&\lstick{\ket{+}} &\qw & \qw& \qw   & \ctrl{2}& \qw& \qw& \qw&\qw\\
&\lstick{\ket{+}} &\qw & \qw& \qw & \qw      & \ctrl{1} &\qw& \qw&\qw\\
&\lstick{\ket{0}} &\qw &\targ & \targ   &\targ  & \targ & \targ& \qw&\qw \\
& &\lstick{\ket{+}} &\ctrl{-1} &\qw&\qw& \qw & \ctrl{-1}&\qw&X\\
}$
}
    }
    \hspace{2em}
    \subfigure[$\ket{\overline{00+}}$ \label{fig:v1-00+}]{
        \resizebox{0.55\linewidth}{!}{
$        
    \Qcircuit @C=0.9em @R=1.1em{
&&\lstick{\ket{0}}&\targ&\qw&\qw&\qw&\targ&\qw&Z\\
&\lstick{\ket{+}} &\qw&\ctrl{-1}& \ctrl{2} & \ctrl{4}& \ctrl{6}&\ctrl{-1}& \qw& \qw&\qw&\qw&\qw& \qw& \qw&\qw&\qw&\qw&\qw
&\gate{Z}&\qw\\
&\lstick{\ket{+}}& \qw &\qw& \qw & \qw& \qw& \ctrl{6}& \qw& \qw&\qw&\qw&\qw& \qw &\qw& \qw & \qw& \qw& \qw &\qw&\qw\\
&\lstick{\ket{0}} & \qw &\qw& \targ & \qw& \qw& \qw& \qw& \qw&\qw&\qw&\qw& \qw&\qw&\qw&\qw&\qw&\qw&\gate{Z}\cwx[-2]&\qw\\
&\lstick{\ket{+}} & \qw &\qw& \qw & \qw& \qw& \qw& \ctrl{4}& \qw&\qw&\qw&\qw&\qw& \qw & \qw& \qw& \qw &\qw&\qw&\qw\\ 
&\lstick{\ket{0}} & \qw &\qw& \qw & \targ& \qw& \qw& \qw& \qw&\qw&\qw&\qw&\qw&\qw&\qw&\qw&\qw&\qw&\gate{Z}\cwx[-2]&\qw\\
&\lstick{\ket{+}}& \qw &\qw & \qw & \qw& \qw& \qw& \qw& \ctrl{2}&\qw&\qw&\qw&\qw& \qw & \qw& \qw& \qw &\qw&\qw&\qw\\
&\lstick{\ket{0}} & \qw &\qw& \qw & \qw& \targ& \qw& \qw& \qw&\qw&\qw&\qw&\qw&\qw&\targ&\qw&\qw&\qw&\gate{Z}\cwx[-2]&\qw\\
&\lstick{\ket{0}} & \qw &\qw& \qw & \qw& \targ& \targ& \targ& \targ&\targ&\qw&\qw&\qw&\targ&\qw& \qw & \qw& \qw& \qw &\qw\\
&&&&&\lstick{\ket{+}}&\ctrl{-1}&\qw&\qw&\qw&\ctrl{-1}&\qw&X\\
& & &&&&&&&&&& &\lstick{\ket{+}}&\ctrl{-2}&\ctrl{-3}&\qw&\gate{H}&\meter &\cctrl{-3}\\
}
$
}
    }
    \subfigure[$\ket{\overline{++0}}$ \label{fig:v1-++0}]{
    \resizebox{0.33\linewidth}{!}{
$        \Qcircuit @C=0.9em @R=1.1em{
            &\lstick{\ket{+}} &\qw & \ctrl{1}&\qw & \qw&\qw& \qw& \qw &\qw &\qw \\
            &\lstick{\ket{0}} &\qw & \targ   &\qw & \qw&\qw& \qw& \qw &\qw &\qw \\
            &\lstick{\ket{+}} &\qw &\qw & \qw& \qw&\qw& \qw& \ctrl{5} &\qw &\qw  \\
            &\lstick{\ket{+}} &\qw &\qw & \qw& \qw&\qw& \ctrl{4}& \qw &\qw &\qw\\ 
            &\lstick{\ket{+}} &\qw &\qw & \qw& \qw& \ctrl{3}&\qw& \qw &\qw &\qw \\
            &\lstick{\ket{+}} &\qw &\qw & \qw& \ctrl{2}& \qw&\qw& \qw &\qw &\qw \\
            &\lstick{\ket{+}} &\qw &\qw & \ctrl{1}& \qw& \qw&\qw& \qw &\qw &\qw \\
            &\lstick{\ket{0}} &\qw &\targ & \targ& \targ& \targ&\targ& \targ &\targ &\qw \\
            & &\lstick{\ket{+}} &\ctrl{-1} &\qw&\qw &\qw &\qw &\qw & \ctrl{-1}&\qw&X\\
}$}
    }
    \hspace{2em}
    \subfigure[$\ket{\overline{{++}+}}$ \label{fig:v1-+++}]{
        \resizebox{0.40\linewidth}{!}{
    $
        \Qcircuit @C=0.9em @R=1.1em{
        &\lstick{\ket{+}} &\qw &\qw & \qw    & \qw& \qw& \qw& \qw& \qw & \ctrl{7}& \qw &\qw\\
        &\lstick{\ket{+}} &\qw &\qw & \qw    & \qw& \qw& \qw& \qw & \ctrl{6}& \qw& \qw &\qw\\
        &\lstick{\ket{+}} &\qw &\qw & \qw   & \qw& \qw& \qw& \ctrl{5}& \qw& \qw& \qw &\qw\\
        &\lstick{\ket{+}} &\qw &\qw & \qw    & \qw& \qw& \ctrl{4}& \qw& \qw& \qw& \qw &\qw\\ 
        &\lstick{\ket{+}} &\qw &\qw & \qw   & \qw& \ctrl{3}& \qw& \qw& \qw& \qw& \qw&\qw\\
        &\lstick{\ket{+}} &\qw &\qw & \qw   & \ctrl{2}& \qw& \qw& \qw& \qw& \qw& \qw&\qw\\
        &\lstick{\ket{+}} &\qw &\qw & \ctrl{1} & \qw      & \qw & \qw& \qw& \qw& \qw& \qw&\qw\\
        &\lstick{\ket{0}} &\qw &\targ & \targ   & \targ      & \targ & \targ& \targ& \targ& \targ& \targ &\qw\\
        & &\lstick{\ket{+}} &\ctrl{-1} &\qw&\qw &\qw &\qw &\qw &\qw &\qw & \ctrl{-1}&\qw&X\\
        }
    $
        }
    }
    \caption{
    Fault-tolerant state-preparation circuits for the Version 1 $[[8,3,2]]$ code: (a) $\ket{\overline{000}}$, (b) $\ket{\overline{00+}}$, (c) $\ket{\overline{++0}}$, and (d) $\ket{\overline{{++}+}}$. Terminal $X$ and $Z$ labels on ancilla lines denote measurements in the corresponding basis. Boxes labeled $X$ or $Z$ connected by classical double lines denote feedforward Pauli corrections conditioned on the measured ancilla outcome.}
    \label{fig:v1-encoding}
\end{figure*}
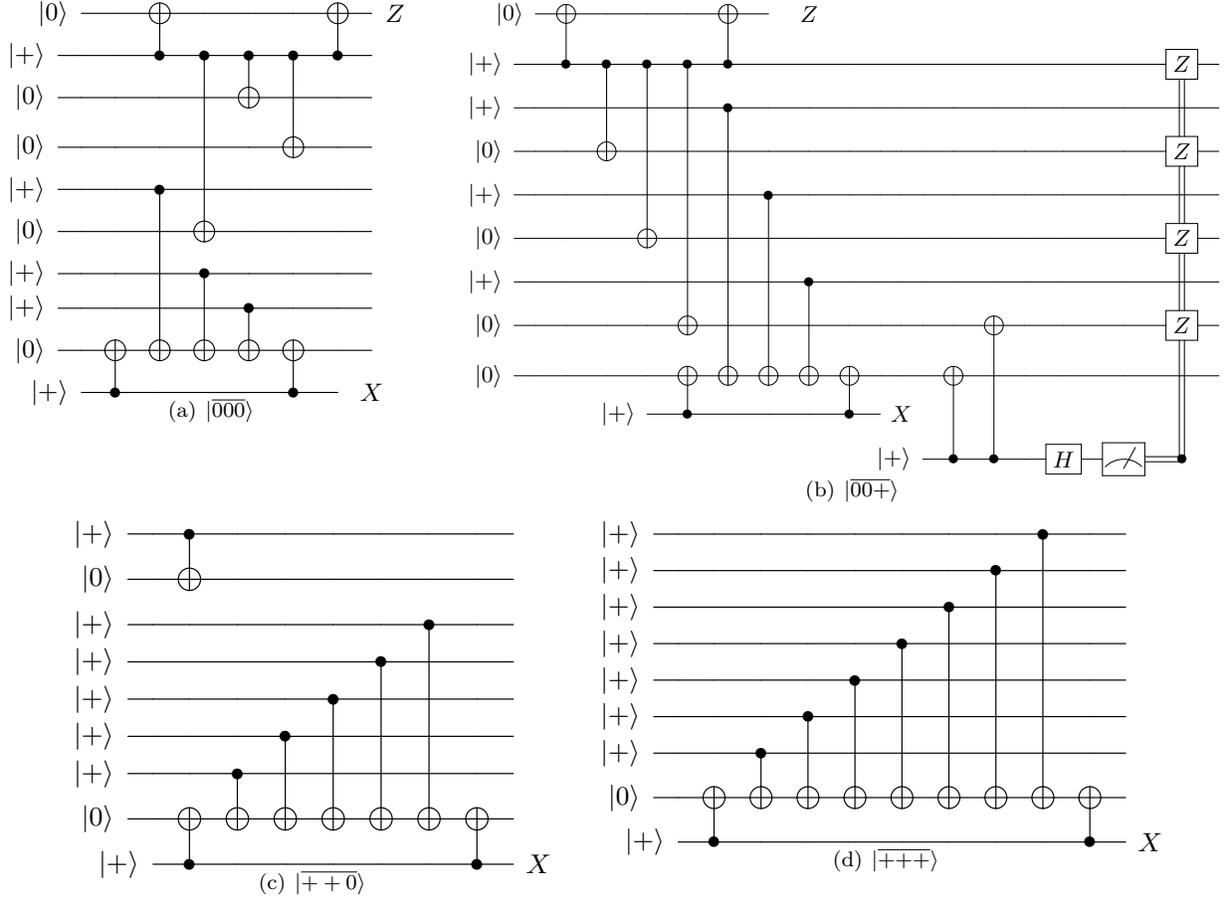

\begin{figure*}[t]
    \centering
    \subfigure[$\ket{\overline{000}}$ \label{fig:v2-000}]{
        \resizebox{0.30\linewidth}{!}{
        $
            \Qcircuit @C=0.9em @R=1.1em{
                &\lstick{\ket{0}} &\targ &\qw&\qw &\qw &\qw &\qw &\qw &\qw & \targ&\qw&Z\\
                &\lstick{\ket{+}} &\ctrl{-1} & \ctrl{1}& \ctrl{2}& \ctrl{3}& \ctrl{4}& \ctrl{5} & \ctrl{6}& \ctrl{7}& \ctrl{-1} &\qw\\
                &\lstick{\ket{0}} &\qw & \targ& \qw& \qw& \qw & \qw&  \qw& \qw& \qw &\qw\\
                &\lstick{\ket{0}} &\qw & \qw& \targ& \qw& \qw& \qw& \qw& \qw& \qw &\qw\\
                &\lstick{\ket{0}} &\qw & \qw& \qw& \targ& \qw& \qw& \qw& \qw& \qw &\qw\\ 
                &\lstick{\ket{0}} &\qw & \qw& \qw& \qw& \targ& \qw& \qw& \qw& \qw &\qw\\ 
                &\lstick{\ket{0}} &\qw & \qw& \qw& \qw& \qw& \targ& \qw& \qw& \qw &\qw\\
                &\lstick{\ket{0}} &\qw & \qw & \qw & \qw& \qw& \qw& \targ& \qw& \qw &\qw\\
                &\lstick{\ket{0}} &\qw &\qw  & \qw & \qw & \qw& \qw& \qw& \targ& \qw &\qw
                }
        $
        }
    }
    \hspace{1em}
    \subfigure[$\ket{\overline{00+}}$ \label{fig:v2-00+}]{
        \resizebox{0.30\linewidth}{!}{
        $
            \Qcircuit @C=0.9em @R=1.1em{
                &&\lstick{\ket{0}}  &\qw &\targ &\qw &\qw &\qw &\qw &\qw &\qw &\targ &\qw &\qw &Z \\
                &&\lstick{\ket{0}} &\targ &\qw &\qw &\qw &\qw &\qw &\qw &\qw &\qw &\targ &\qw &Z \\
                &\lstick{\ket{+}} &\qw &\ctrl{-1} &\qw & \ctrl{2}& \qw& \ctrl{4}& \qw& \ctrl{6} & \qw &\qw &\ctrl{-1} &\qw \\
                &\lstick{\ket{+}} &\qw &\qw &\ctrl{-3} & \qw& \ctrl{2}& \qw& \ctrl{4}& \qw& \ctrl{6} & \ctrl{-3} &\qw &\qw \\
                &\lstick{\ket{0}} &\qw &\qw &\qw & \targ& \qw& \qw& \qw& \qw& \qw & \qw &\qw &\qw\\
                &\lstick{\ket{0}} &\qw &\qw &\qw & \qw& \targ& \qw& \qw& \qw& \qw & \qw &\qw &\qw\\ 
                &\lstick{\ket{0}} &\qw &\qw &\qw & \qw& \qw& \targ& \qw& \qw& \qw & \qw &\qw &\qw\\ 
                &\lstick{\ket{0}} &\qw &\qw &\qw & \qw& \qw& \qw& \targ& \qw& \qw & \qw &\qw &\qw\\
                &\lstick{\ket{0}} &\qw &\qw &\qw & \qw& \qw& \qw& \qw& \targ& \qw & \qw &\qw &\qw\\
                &\lstick{\ket{0}} &\qw &\qw &\qw & \qw& \qw& \qw& \qw& \qw& \targ & \qw &\qw &\qw
                }
        $
        }
    }
    \hspace{1em}
    \subfigure[$\ket{\overline{++0}}$ \label{fig:v2-++0}]{
        \resizebox{0.30\linewidth}{!}{
        $
            \Qcircuit @C=0.9em @R=1.1em{
                &\lstick{\ket{+}}& \ctrl{1}&\ctrl{8}&\qw&\qw&\qw&\qw&\qw&\qw&\qw&\qw&\qw\\
                &\lstick{\ket{0}}& \targ&\qw&\qw&\qw&\qw&\qw&\qw&\qw&\qw&\qw&\qw\\
                &\lstick{\ket{+}}& \ctrl{1}&\qw&\qw&\ctrl{6}&\qw&\qw&\qw&\qw&\qw&\qw&\qw\\
                &\lstick{\ket{0}}& \targ&\qw&\qw&\qw&\qw&\qw&\qw&\qw&\qw&\qw&\qw\\ 
                &\lstick{\ket{+}}& \ctrl{1}&\qw&\qw&\qw&\ctrl{4}&\qw&\qw&\qw&\qw&\qw&\qw\\ 
                &\lstick{\ket{0}}& \targ&\qw&\qw&\qw&\qw&\qw&\qw&\qw&\qw&\qw&\qw\\
                &\lstick{\ket{+}}& \ctrl{1}&\qw&\qw&\qw&\qw&\qw&\ctrl{2}&\qw&\gate{X}&\qw\\
                &\lstick{\ket{0}}& \targ&\qw&\qw&\qw&\qw&\qw&\qw&\qw&\gate{X}\cwx[-1]&\qw\\
                &\lstick{\ket{0}}&\qw&\targ&\targ&\targ&\targ&\targ&\targ&\meter &\cctrl{-1}\\
                &&&\lstick{\ket{+}}&\ctrl{-1} &\qw&\qw&\ctrl{-1}&\qw&X
                }
        $
        }
    }
    \subfigure[$\ket{\overline{{++}+}}$ from \cite{honciuc2024implementing}\label{fig:v2-+++}]{
        \includegraphics[width=0.6\textwidth]{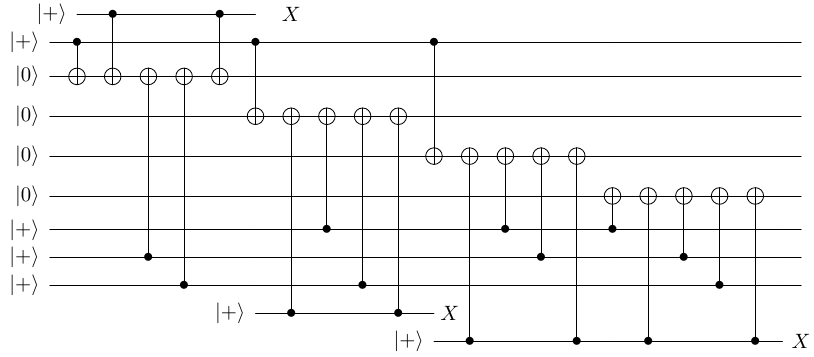}
    }
    \caption{
    Fault-tolerant state-preparation circuits for the Version 2 $[[8,3,2]]$ code: (a) $\ket{\overline{000}}$, (b) $\ket{\overline{00+}}$, (c) $\ket{\overline{++0}}$, and (d) $\ket{\overline{{++}+}}$ (from Ref.~\cite{honciuc2024implementing}). Terminal $X$ and $Z$ labels on ancilla lines denote measurements in the corresponding basis. Boxes labeled $X$ or $Z$ connected by classical double lines denote feedforward Pauli corrections conditioned on the measured ancilla outcome.}
    \label{fig:v2-encoding}
\end{figure*}

\subsection{Version 1}

The Version 1 preparation circuits are collected in \cref{fig:v1-encoding}.

\paragraph{Logical $\ket{\overline{000}}$ (\cref{fig:v1-000}).}
This preparation fixes all three logical qubits of the parent $[[8,3,3,2]]$ subsystem code to $\ket{\overline{0}}$ and all three gauge qubits to $\ket{+_g}$. Since all relevant logical and gauge operators are weight-$2$, we can reuse the fault-tolerant preparation of logical $\ket{\overline{+^j0^{n-j-2}}}$ states in the $[[n,n-2,2]]$ code from \cite{chao2018quantum}, after changing basis. In particular, the required state is equivalent to the logical $\ket{\overline{+^3 0^3}}$ state of the $[[8,6,2]]$ stabilizer code with logical operator basis
\beq
    X_k X_7,\quad Z_0 Z_k,\qquad k=1,\dots,6.
\eeq
Because $j=3$ is odd, this state factorizes into a four-qubit dual GHZ state and a four-qubit GHZ state:
\beq
    \begin{aligned}
        \ket{\overline{+^3 0^3}}
        =&\tfrac{1}{\sqrt{2}}
        \left(\ket{+_3 +_5 +_6 +_7}+\ket{-_3 -_5 -_6 -_7}\right) \\
        \otimes&\tfrac{1}{\sqrt{2}}
        \left(\ket{0_0 0_1 0_2 0_4}+\ket{1_0 1_1 1_2 1_4}\right).
    \end{aligned}
\eeq
This gives the Version 1 logical $\ket{\overline{000}}$ state.

\paragraph{Logical $\ket{\overline{{++}+}}$ (\cref{fig:v1-+++}).}
Here, all three logical qubits and all three gauge qubits are fixed to $\ket{\overline{+}}$ and $\ket{+_g}$, respectively. The $+1$ eigenoperators reduce as
\beq
    \begin{aligned}
        & \overline{X}_1=X_0 X_1 X_2 X_3 \to X_6 X_4 \\
        & \overline{X}_2=X_0 X_1 X_4 X_5 \to X_6 X_2 \\
        & \overline{X}_3=X_0 X_2 X_4 X_6 \to X_6 X_1 \\
        & G^X_4=X_7 X_3 \to X_3 X_6 \\
        & G^X_5=X_7 X_5 \to X_5 X_6 \\
        & G^X_6=X_7 X_6 \to X_7 X_6 \\
        & S^X=X^{\otimes 8} \to X_0 X_6 \\
        & S^Z=Z^{\otimes 8} \to Z^{\otimes 8}.
    \end{aligned}
\eeq
These reduced generators are those of the $8$-qubit dual GHZ state, so the preparation reduces directly to the dual-GHZ gadget.

\paragraph{Logical $\ket{\overline{++0}}$ (\cref{fig:v1-++0}).}
This is the native mixed-basis state in Version 1. Using $\ket{\overline{++0}}$ as a representative, the Gaussian-eliminated generators are identical to those of $\ket{\overline{{++}+}}$ except that $X^{\otimes 8}$ is replaced by $X_0 X_1$. Together with the additional operator $Z_0 Z_1$, this isolates a Bell pair on qubits $0$ and $1$ and a six-qubit dual GHZ state on the remaining qubits:
\beq
    \begin{aligned}
        \ket{\overline{++0}}
        =&\tfrac{1}{\sqrt{2}}
        \left(\ket{+_3 +_5 +_6 +_7 +_2 +_4}+\ket{-_3 -_5 -_6 -_7 -_2 -_4}\right) \\
        \otimes&\tfrac{1}{\sqrt{2}}
        \left(\ket{0_0 0_1}+\ket{1_0 1_1}\right).
    \end{aligned}
\eeq
The Bell state is prepared fault-tolerantly without ancillas, and the six-qubit dual GHZ state is prepared fault-tolerantly with one ancilla. Other orderings of $\ket{\overline{+^20^1}}$ are obtained analogously.

\paragraph{Logical $\ket{\overline{00+}}$ (\cref{fig:v1-00+}).}
This is the non-native mixed-basis state in Version 1, so we prepare it through a partially gauge-fixed intermediate state. We first fix $G^X_4$, $G^X_5$, and $G^Z_6$. For the representative state $\ket{\overline{00+}}$, the $+1$ eigenoperators reduce to
\beq
    \begin{aligned}
        & \overline{Z}_1 = Z_0 Z_4 \to Z_0 Z_4 \\
        & \overline{Z}_2 = Z_0 Z_2 \to Z_0 Z_2 \\
        & \overline{X}_3 = X_0 X_2 X_4 X_6 \to X_1 X_7 \\
        & G^X_4 = X_7 X_3 \to X_7 X_3 \\
        & G^X_5 = X_7 X_5 \to X_7 X_5 \\
        & G^Z_6 = Z_0 Z_2 Z_4 Z_6 \to Z_0 Z_6 \\
        & S^X = X^{\otimes 8} \to X_0 X_2 X_4 X_6 \\
        & S^Z = Z^{\otimes 8} \to Z_1 Z_3 Z_5 Z_7.
    \end{aligned}
\eeq
These reduced generators describe a four-qubit GHZ state on the even qubits $\{0,2,4,6\}$ and a four-qubit dual GHZ state on the odd qubits $\{1,3,5,7\}$. After preparing this intermediate state, we measure the single remaining gauge operator
$G^X_6 = X_7 X_6$
together with its complementary check
$X_0 X_1 X_2 X_3 X_4 X_5$,
whose product gives the pre-correction $X^{\otimes 8}$ syndrome. Because only one gauge operator remains to be measured, there are no later gauge measurements whose outcomes can be corrupted by an intermediate fault, so an additional dedicated $X^{\otimes 8}$ stabilizer measurement is not needed inside this subroutine. Any residual data error is handled by the stabilizer and gauge checks immediately after state preparation.

\subsection{Version 2}

The Version 2 preparation circuits are collected in \cref{fig:v2-encoding}.

\paragraph{Logical $\ket{\overline{000}}$ (\cref{fig:v2-000}).}
This preparation fixes all three logical qubits and all three gauge qubits of the parent $[[8,3,3,2]]$ subsystem code to $\ket{\overline{0}}$ and $\ket{0_g}$, respectively. The $+1$ eigenoperators reduce as
\beq
    \begin{aligned}
        & \overline{Z}_1=Z_0 Z_4 \to Z_0 Z_4 \\
        & \overline{Z}_2=Z_0 Z_2 \to Z_0 Z_2 \\
        & \overline{Z}_3=Z_0 Z_1 \to Z_0 Z_1 \\
        & G^Z_4=Z_0 Z_1 Z_2 Z_3 \to Z_0 Z_3 \\
        & G^Z_5=Z_0 Z_1 Z_4 Z_5 \to Z_0 Z_5 \\
        & G^Z_6=Z_0 Z_2 Z_4 Z_6 \to Z_0 Z_6 \\
        & S^X=X^{\otimes 8} \to X^{\otimes 8} \\
        & S^Z=Z^{\otimes 8} \to Z_0 Z_7.
    \end{aligned}
\eeq
These are the generators of the $8$-qubit GHZ state, so the logical $\ket{\overline{000}}$ preparation reduces to the GHZ gadget.

\paragraph{Logical $\ket{\overline{{++}+}}$ (\cref{fig:v2-+++}).}
For Version 2, we use the fault-tolerant preparation circuit for $\ket{\overline{{++}+}}$ introduced in Ref.~\cite{honciuc2024implementing}.

\paragraph{Logical $\ket{\overline{00+}}$ (\cref{fig:v2-00+}).}
This is the native mixed-basis state in Version 2. Using $\ket{\overline{00+}}$ as a representative, the $+1$ eigenoperators reduce to
\beq
    \begin{aligned}
        & \overline{Z}_1=Z_0 Z_4 \to Z_0 Z_4 \\
        & \overline{Z}_2=Z_0 Z_2 \to Z_0 Z_2 \\
        & \overline{X}_3=X_0 X_2 X_4 X_6 \to X_0 X_2 X_4 X_6 \\
        & G^Z_4=Z_0 Z_1 Z_2 Z_3 \to Z_1 Z_3 \\
        & G^Z_5=Z_0 Z_1 Z_4 Z_5 \to Z_1 Z_5 \\
        & G^Z_6=Z_0 Z_2 Z_4 Z_6 \to Z_0 Z_6 \\
        & S^X=X^{\otimes 8} \to X_1 X_3 X_5 X_7 \\
        & S^Z=Z^{\otimes 8} \to Z_1 Z_7.
    \end{aligned}
\eeq
These reduced generators describe two independent four-qubit GHZ states, one on the even qubits $\{0,2,4,6\}$ and one on the odd qubits $\{1,3,5,7\}$. The logical state is therefore prepared by fault-tolerantly preparing those two GHZ states.

\paragraph{Logical $\ket{\overline{++0}}$ (\cref{fig:v2-++0}).}
This is the non-native mixed-basis state in Version 2, so again we use a partially gauge-fixed intermediate state. We first fix $G^Z_4$, $G^Z_5$, and $G^X_6$. For the representative state $\ket{\overline{++0}}$, the reduced $+1$ eigenoperators are
\beq
    \begin{aligned}
        & \overline{X}_1 = X_0 X_1 X_2 X_3 \to X_4 X_5 \\
        & \overline{X}_2 = X_0 X_1 X_4 X_5 \to X_0 X_1 \\
        & \overline{Z}_3 = Z_0 Z_1 \to Z_0 Z_1 \\
        & G^Z_4 = Z_0 Z_1 Z_2 Z_3 \to Z_2 Z_3 \\
        & G^Z_5 = Z_0 Z_1 Z_4 Z_5 \to Z_4 Z_5 \\
        & G^X_6 = X_7 X_6 \to X_7 X_6 \\
        & S^X = X^{\otimes 8} \to X_2 X_3 \\
        & S^Z = Z^{\otimes 8} \to Z_7 Z_6.
    \end{aligned}
\eeq
These generators define four Bell states on the qubit pairs $\{0,1\}$, $\{2,3\}$, $\{4,5\}$, and $\{6,7\}$, so the intermediate state is prepared fault-tolerantly without ancillas. We then measure the single remaining gauge operator
$G^Z_6 = Z_0 Z_2 Z_4 Z_6$
together with its complementary check
$Z_1 Z_3 Z_5 Z_7$,
whose product gives the pre-correction $Z^{\otimes 8}$ syndrome. As in the Version 1 construction above, no additional dedicated $Z^{\otimes 8}$ stabilizer measurement is required inside this single-step gauge-fixing subroutine, because there are no later gauge measurements whose outcomes can be corrupted by an intermediate fault. Any residual data error is handled by the stabilizer and gauge checks applied immediately after state preparation.

\subsection{Remarks on circuit simplifications}
Single-qubit errors may propagate to higher-weight errors through an encoding circuit. If these propagated errors can be detected by measuring the ancillas and stabilizers in the following subroutines, the encoding circuit is fault-tolerant. \Cref{fig:sim_v12_init} has shown that the encoding circuits in this section are fault-tolerant when the corresponding stabilizers and gauge operators are measured immediately after the encoding circuits. 

Measuring the $X^{\otimes 8}$ and $Z^{\otimes 8}$ stabilizers is sufficient to verify that the state lies in the stabilizer code space.
However, as discussed in \cref{sec:4}, this is not always sufficient to guarantee the correct gauge sector for subsequent logical operations, because incorrectly applied gauge operators can act as logical faults.
Some encoding circuits therefore do not require any gauge measurement, whereas others do. We discuss the different scenarios below. Note that fault-tolerant measurement of weight-4 gauge operators requires flag qubits \cite{chao2018quantum}, as shown in \cref{fig:measure}.

\paragraph{All the native encoding circuits with GHZ states.} They do not need to measure any gauge operator immediately after, because all the propagated errors act as stabilizers. 

\paragraph{All the non-native encoding circuits.} A single Pauli error on the ancilla before measuring the gauge operator can change the measurement result and result in an incorrectly applied gauge operator. Hence, the same gauge operator must be measured again to align with the first measurement result. For example, when the $\ket{\overline{00+}}$ state in Version 1 is prepared, an $X_0$ error on the ancilla to measure the $G^X_6$ gauge operator can cause the incorrect application of the $G^Z_6$ gauge operator. The $G^X_6$ gauge operator must be measured again to compare with the initial measurement.

\paragraph{The logical $\ket{\overline{{++}+}}$ state in Version 2.}
In this fault-tolerant preparation circuit, the propagated errors $X_0X_3$, $X_4X_5$, $X_4X_6$, and $X_4X_7$ are not detectable by measuring only the $Z^{\otimes 8}$ stabilizer. Moreover,
\beq
\begin{aligned}
X_0X_3&=\overline{X}_1\overline{X}_2\overline{X}_3\,G^X_5G^X_6 \\
X_4X_5&=X^{\otimes 8}\overline{X}_1G^X_6\\
X_4X_6&=X^{\otimes 8}\overline{X}_1G^X_5\\
X_4X_7&=X^{\otimes 8}\overline{X}_1G^X_5G^X_6\\
\end{aligned}
\eeq
Therefore, on the logical state $\ket{\overline{{++}+}}$, for which $\overline{X}_1=\overline{X}_2=\overline{X}_3=+1$ and $X^{\otimes 8}=+1$, these propagated errors are equivalent to the gauge operators $G^X_5$, $G^X_6$, or $G^X_5G^X_6$. As in \cref{sec:4}, incorrectly applied gauge operators can cause logical errors
under subsequent Version~2 logical gates, so a standalone verification of this preparation circuit requires measuring both $G^Z_5$ and $G^Z_6$. 
This verification strategy is 
used in the Grover algorithm simulation in \cref{sec:Grover}.

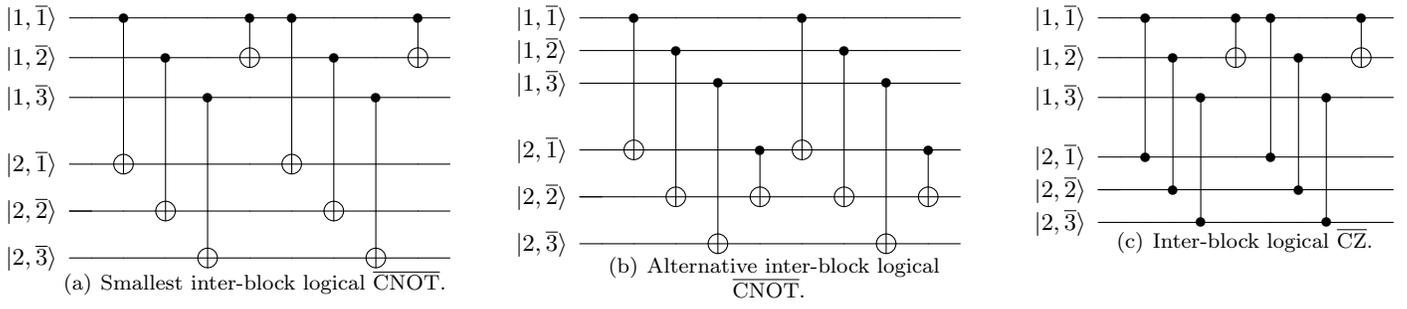
\begin{figure*}[ht]
    \centering
    \subfigure[Smallest inter-block logical $\overline{\mathrm{CNOT}}$.
        \label{fig:smallest-interblock-cnot}]{
        $\Qcircuit @C=0.9em @R=1.1em{
        \lstick{\ket{1,\overline{1}}}& \qw &\ctrl{4} &\qw & \qw&\ctrl{1}&\ctrl{4} &\qw & \qw&\ctrl{1}&\qw\\
        \lstick{\ket{1,\overline{2}}}& \qw &\qw  &\ctrl{4} &\qw&\targ&\qw  &\ctrl{4} &\qw&\targ&\qw\\
        \lstick{\ket{1,\overline{3}}}& \qw &\qw  &\qw &\ctrl{4} &\qw&\qw  &\qw &\ctrl{4}&\qw&\qw\\
        \\
        \lstick{\ket{2,\overline{1}}}& \qw&\targ&\qw&\qw&\qw&\targ&\qw&\qw& \qw&\qw\\
        \lstick{\ket{2,\overline{2}}}& \qw\qw &\qw&\targ&\qw&\qw&\qw&\targ&\qw& \qw&\qw\\
        \lstick{\ket{2,\overline{3}}}& \qw &\qw&\qw&\targ&\qw&\qw&\qw&\targ&\qw&\qw\\
        }$
    }
    \hfill
    \subfigure[Alternative inter-block logical $\overline{\mathrm{CNOT}}$.
        \label{fig:alt-interblock-cnot}]{
        $\Qcircuit @C=0.9em @R=1.1em{
        \lstick{\ket{1,\overline{1}}}& \qw &\ctrl{4} &\qw & \qw&\qw&\ctrl{4} &\qw & \qw&\qw&\qw\\
        \lstick{\ket{1,\overline{2}}}& \qw &\qw  &\ctrl{4} &\qw&\qw&\qw  &\ctrl{4} &\qw&\qw&\qw\\
        \lstick{\ket{1,\overline{3}}}& \qw &\qw  &\qw &\ctrl{4} &\qw&\qw  &\qw &\ctrl{4}&\qw&\qw\\
        \\
        \lstick{\ket{2,\overline{1}}}& \qw&\targ&\qw&\qw&\ctrl{1}&\targ&\qw&\qw& \ctrl{1}&\qw\\
        \lstick{\ket{2,\overline{2}}}& \qw\qw &\qw&\targ&\qw&\targ&\qw&\targ&\qw& \targ&\qw\\
        \lstick{\ket{2,\overline{3}}}& \qw &\qw&\qw&\targ&\qw&\qw&\qw&\targ&\qw&\qw\\
        }$
        \label{fig:alt-interblock-cnot}
    }
    \hfill
    \subfigure[Inter-block logical $\overline{\mathrm{CZ}}$.
        \label{fig:interblock-cz}]{
        $\Qcircuit @C=0.9em @R=1.1em{
        \lstick{\ket{1,\overline{1}}}& \qw &\ctrl{4} &\qw & \qw&\ctrl{1}&\ctrl{4} &\qw & \qw&\ctrl{1}&\qw\\
        \lstick{\ket{1,\overline{2}}}& \qw &\qw  &\ctrl{4} &\qw&\targ&\qw  &\ctrl{4} &\qw&\targ&\qw\\
        \lstick{\ket{1,\overline{3}}}& \qw &\qw  &\qw &\ctrl{4} &\qw&\qw  &\qw &\ctrl{4}&\qw&\qw\\
        \\
        \lstick{\ket{2,\overline{1}}}& \qw&\ctrl{0}&\qw&\qw&\qw&\ctrl{0}&\qw&\qw& \qw&\qw\\
        \lstick{\ket{2,\overline{2}}}& \qw\qw &\qw&\ctrl{0}&\qw&\qw&\qw&\ctrl{0}&\qw& \qw&\qw\\
        \lstick{\ket{2,\overline{3}}}& \qw &\qw&\qw&\ctrl{0}&\qw&\qw&\qw&\ctrl{0}&\qw&\qw\\
        }$
        \label{fig:interblock-cz}
    }

    \label{fig:interblock-gates}
    \caption{Inter-block logical gates between blocks 1 and 2.}
\end{figure*}

\section{Extensions of the single inter-block logical $\overline{\mathrm{CNOT}}$ gate}
\label{app:C}

The logical $\overline{\mathrm{SWAP}}$ gates in \cref{fig:interblock-CNOT} are used only to move the target from $\ket{2,\overline{2}}$ to $\ket{2,\overline{1}}$; they are not required for the core construction. Removing them yields \cref{fig:smallest-interblock-cnot}, which implements a single inter-block logical $\overline{\mathrm{CNOT}}$ from logical qubit $\overline{1}$ of block $1$ to logical qubit $\overline{2}$ of block $2$.

More generally, two layers of parallel inter-block logical $\overline{\mathrm{CNOT}}$ gates, separated by an intra-block logical $\overline{\mathrm{CNOT}}(\overline{i}\to\overline{j})$ on one block and followed by the same intra-block gate again, cancel the unwanted pairwise couplings and leave only the desired single inter-block logical $\overline{\mathrm{CNOT}}$ from logical qubit $\overline{i}$ in one block to logical qubit $\overline{j}$ in the other. In \cref{fig:smallest-interblock-cnot} the intra-block gate is applied on block $1$, whereas \cref{fig:alt-interblock-cnot} shows the equivalent construction with the intra-block gate applied on block $2$. The latter form is useful when only one of the two blocks admits a convenient fault-tolerant intra-block logical $\overline{\mathrm{CNOT}}$ implementation.

The same cancellation idea also gives a single inter-block logical $\overline{\mathrm{CZ}}$: replacing the two layers of parallel inter-block $\overline{\mathrm{CNOT}}$ gates by layers of parallel inter-block $\overline{\mathrm{CZ}}$ gates yields the construction shown in \cref{fig:interblock-cz}.

\end{document}